\documentclass{winnower}
\usepackage{amsmath,amssymb}
\usepackage{booktabs}
\usepackage{graphics}
\usepackage{graphicx}
\usepackage{mathdots,mathrsfs}
\usepackage{moresize}
\usepackage{multicol}
\usepackage{multirow}
\usepackage{natbib}
\usepackage{setspace}
\usepackage{caption}
\usepackage[table,xcdraw]{xcolor}
\usepackage{subfigure}
\usepackage{minipage}
\usepackage{fancyhdr}
\usepackage{yhmath}
\newcommand{\er}{Erd\H{o}s-R\'{e}nyi }
\usepackage{wrapfig}
\usepackage{framed}
\usepackage{sidecap}
\usepackage{mwe}
\usepackage{rotating}
\usepackage{adjustbox}
\usepackage{blindtext}
\usepackage{tikz}
\usepackage{longtable}
\usepackage{pdfpages}
\usepackage{array}
\newcolumntype{R}[1]{>{\raggedright\let\newline\\\arraybackslash\hspace{0pt}}m{#1}}
\begin{document}

\title{Generalizations of Edge Overlap to Weighted and Directed Networks}

\author{Heather Mattie}
\author{Jukka-Pekka Onnela}
\affil{Department of Biostatistics, Harvard T.H. Chan School of Public Health, Boston, MA}

\date{}

\maketitle

\begin{abstract} 
With the increasing availability of behavioral data from diverse digital sources, such as social media sites and cell phones, it is now possible to obtain detailed information about the structure, strength, and directionality of social interactions in varied settings. While most metrics of network structure have traditionally been defined for unweighted and undirected networks only, the richness of current network data calls for extending these metrics to weighted and directed networks. One fundamental metric in social networks is edge overlap, the proportion of friends shared by two connected individuals. Here we extend definitions of edge overlap to weighted and directed networks, and present closed-form expressions for the mean and variance of each version for the \er random graph and its weighted and directed counterparts. We apply these results to social network data collected in rural villages in southern Karnataka, India. We use our analytical results to quantify the extent to which the average overlap of the empirical social network deviates from that of corresponding random graphs and compare the values of overlap across networks. Our novel definitions allow the calculation of edge overlap for more complex networks and our derivations provide a statistically rigorous way for comparing edge overlap across networks. 
 \end{abstract}

\section{Introduction}

\par Humans interact with each other both online and in-person, forming and dissolving social ties throughout our lives. The flexible architecture of networks, or graphs, make them a useful paradigm for modeling these complex relationships at the individual, group, and population levels. Social network nodes typically represent individuals and edges the connections between individuals, such as friendships, sexual contacts, or cell phone calls. Social networks have been shown to have a direct impact on public health \cite{Chris,Chris2,Fowl,Fowl2,Good}. 

For example, a recent study on polio examined the social networks of households in Malegaon, India. Each household was assigned to one of three categories based on their polio vaccine status: pro-vaccine (accepted the vaccine the first time), vaccine-reluctant (initially resisted but eventually accepted the vaccine), and vaccine-refusing (never accepted the vaccine). Vaccine-refusing households had a greater proportion of their ties to vaccine-reluctant and vaccine-refusing households than pro-vaccine households did, indicating clustering of households by their vaccine status \cite{Polio}. Several studies have now successfully modeled the spread of epidemics through various populations, finding that different network structures have an effect on the potential efficacy of an intervention \cite{Banerjee, Valente, VanderWeele}. Studies have also leveraged network properties to target highly connected individuals in public health interventions \cite{Kim2}. The structure of connections in contact networks have also been shown to affect statistical power in cluster randomized trials \cite{Banerjee, Staples}. Additionally, new classes of connectivity-informed study designs for cluster randomized trials have been proposed recently, and these designs can be used to achieve the dual goal of studying the effect of the proposed new intervention and controlling the epidemic simultaneously \cite{Banerjee, Kim, Harling}. There is also accumulating evidence that the habits of our friends influence our own behavior, such as the uptake of smoking or lifestyle choices that can lead to obesity \cite{Chris, Chris2, Fowl, Fowl2}. Moreover, electronic billing records have been used to study patient-physician interaction networks to learn about structural properties of these networks and how these properties are associated with the quality and cost of health care \cite{Landon, Kim, Sima}.

All of the above studies and results showcase the influence network structure has on human behavior and the spread of simple (one source) and complex (multiple sources) contagion. Network structure can be studied at different scales ranging from local to global. Microscopic (local) structures include one or a few nodes, macroscopic (global) structures involve most to all nodes, and mesoscopic structures lie between the two extremes. It has been shown that the different structures are not independent of one another \cite{Fort}. Specifically, several combinations of microscopic mechanisms are known to give rise to unexpected mesoscopic and macroscopic structure \cite{Bianconi, Fort, Kumpula}. For example, triadic closure, the process of getting to know a friend of a friend, can generate network communities \cite{Kumpula, Porter}. The term community here refers to a group of nodes that are densely connected to one another but only sparsely connected to the rest of the network \cite{Porter, Kumpula}. Community structure is of particular interest because most social networks have meaningful community structure that is related to their function. Communities also arise from humans forming tightly-knit groups through shared interests and similar characteristics, and they play an important role in the spread of disease and information, and the adoption of behaviors \cite{Chris, Chris2, Fort}. 

There are many ways and many levels for characterizing the notion of clustering in a network. Starting from the macroscopic view, connected components can be seen as clusters within a network, such that each node in the component is reachable (via a finite path) from any other node in the component. Network communities impose in some sense a stricter requirement of connectivity. Although there are many ways to formulate network communities, a common requirement is that the nodes within a community should not only be reachable from other nodes in the community but in addition the nodes should be densely connected with one another. From the microscopic perspective, node clustering refers to the number of edges among the neighbors of a given node divided by the maximum number of such edges that could exist, and various ways of formalizing this notion exist \cite{Watts, Saramaki}. Edge clustering is similar to node clustering, and it usually quantifies the proportion of ``triangles'' (strictly, closed triples) around an edge, i.e., the number of triangles that use the given edge as one of the edges of the triangle, divided by the maximum number of such triangles that could exist. The normalization term here usually requires taking the minimum of two terms, which makes it analytically more difficult to deal with \cite{Wang}. Edge overlap, the focus of this paper, is very closely related to edge clustering, but its formulation is somewhat different and it is more closely motivated by the concept of triadic closure and the weak ties hypothesis of Granovetter \cite{Granovetter}.

Edge overlap is defined as the proportion of common friends two connected individuals share among all of their friends \cite{Onnela, Chou}. While higher overlap increases the spread of simple contagions within communities, it plays a particularly important part in complex contagion. Here, adopting a new behavior, idea or innovation requires exposure to several sources, possibly several times before adoption or contagion can occur. Concrete examples include adopting a new diet or political views if most of your friendship circle has adopted them first. Because complex contagions require a great amount of social reinforcement, and individuals typically befriend those similar to themselves, higher levels of edge overlap provide more sources of a particular exposure and speed the spread of these contagions \cite{Centola}.

Social network data has traditionally been collected from surveys, mostly capturing small, static network snapshots at one point in time \cite{Faust}. Dozens of different metrics, including edge overlap, have been created to quantify and study the structure of these networks. However, with the recent availability of increasingly rich, complex network data, limitations of some of these metrics have become increasingly clear. For example, betweenness centrality, the proportion of all pairwise shortest paths in a network that pass through a specified node, is used quite broadly to quantify the centrality or importance of a node, but it becomes more computationally demanding as the size of the network increases and, even more importantly, its interpretation or usefulness in very large social networks is less clear. Another example of a widely used metric is the clustering coefficient, which has subsequently been extended to weighted and directed networks \cite{Saramaki, Tore}. Here we define extensions of edge overlap, currently only defined for unweighted and undirected networks, for weighted and directed networks. 

The rest of this paper is organized as follows. In Section 2 we introduce edge overlap and define extensions of edge overlap for weighted and directed networks. We then present two closed-form expressions for the mean and variance of edge overlap for the \er random graph and its weighted and directed counterparts. We then demonstrate the accuracy of our mean and variance approximations through simulation in Section 3. We apply our results to social network data collected in southern Karnataka, India and quantify the difference in average overlap for this empirical network to the expected average overlap for a corresponding random graph in Section 4. In Section 5 we present the results of our data analysis and discuss our results and conclusions in Section 6. Supplementary material including mean and variance derivations is contained in Appendices A and B.


\section{Methods}

\subsection{Overlap Extensions}
\begin{figure}[h]
\centering
\includegraphics[scale = 0.6]{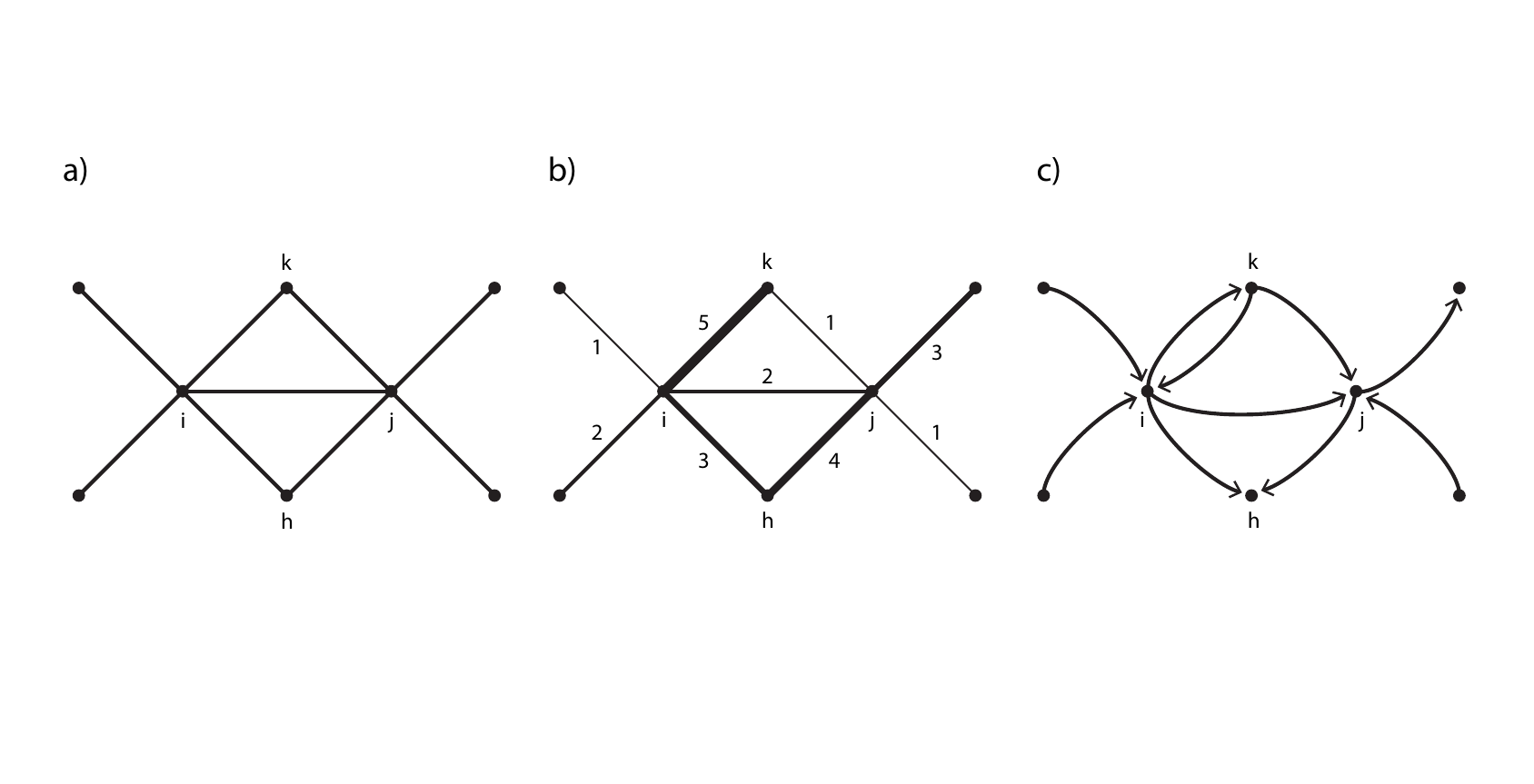}
\vspace{-55pt}
\captionsetup{width=\textwidth}
\caption{Schematics of edge overlap for (a) an unweighted network, (b) weighted network, and (c) directed (unweighted) network. Nodes are labeled with letters and weights are labeled with numbers.}
   \label{fig:schematic}
\end{figure}

An important microscopic metric, which captures the overlap of the friendship circles for two connected individuals, is the edge overlap \cite{Onnela}. In mathematical terms, the overlap between two connected nodes $i$ and $j$ is defined as
\begin{equation}\label{eq:1}
o_{ij} = \frac{n_{ij}}{(k_i -1) + (k_j -1) - n_{ij}}
\end{equation}
where $n_{ij}$ is the number of common neighbors of nodes $i$ and $j$, and $k_i$ $(k_j)$ denotes the degree, or number of connections, node $i$ $(j)$ has. Note that the tie between nodes $i$ and $j$ is not included in the calculation; overlap for the edge $(i, j)$ is defined only where $A_{ij} = 1$ and $k_1 + k_j >2$. As a concrete example, suppose we want to calculate edge overlap for nodes $i$ and $j$ in an undirected network as shown in Figure \ref{fig:schematic}a. Here, $i$ has a total of five friends, $j$ has a total of five friends, and $i$ and $j$ have two friends in common: nodes $h$ and $k$. Thus, the value of edge overlap can be calculated as $o_{ij} = n_{ij}/((k_i -1) + (k_j -1) - n_{ij}) = 2/((5-1) + (5-1) - 2) = 1/3$.

Currently, edge overlap is only defined for simple networks consisting of unweighted and undirected edges \cite{Onnela}. Moreover, expressions for the mean and variance of edge overlap for a particular network do not yet exist, making it difficult to carry out principled comparisons of this metric across networks, in particular, networks of different sizes.

In a weighted network, each edge has a weight assigned to it. We define weighted overlap in Eq. \eqref{eq:2} as the proportion of weight associated with ties that are adjacent to common neighbors of nodes $i$ and $j$:
\begin{eqnarray}\label{eq:2}
o^W_{ij} = \frac{\sum^{n_{ij}}_{k=1}(w_{ik} + w_{jk})}{s_i + s_j - 2w_{ij}}.
\end{eqnarray}
Here, $n_{ij}$ is the number of common neighbors of nodes $i$ and $j$, $w_{ij}$ denotes the weight associated with the tie between nodes $i$ and $j$, and $s_i$ $(s_j)$ denotes the strength of node $i$ $(j)$, i.e., the sum of all edge weights associated with node $i$ ($j$). According to the definition, we first identify the common neighbors of nodes $i$ and $j$, then sum together the weights associated with these edges, and finally divide this sum by the combined strengths of nodes $i$ and $j$, where we exclude the tie that connects the two nodes from the computation of each node's strength. The last step is intended to ensure consistency with the original version of edge overlap, i.e., the weight of the tie between the two individuals being considered is not included in the calculation of $o^W_{ij} $. Also, the metric is only defined for $w_{ij} > 0$ and for $s_i + s_j > 2w_{ij}$. For nodes $i$ and $j$ in Figure \ref{fig:schematic}b, the weighted overlap can be calculated as $o^W_{ij} = \sum^{n_{ij}}_{k=1}(w_{ik} + w_{jk}) / (s_i + s_j - 2w_{ij}) = ((5+1)+(3+4)) / (13 + 11 - 4) = 13 / 20$.

One subtle but important detail to point out about our weighted overlap definition is that $o^W_{ij}$ does not reduce to $o_{ij}$ when all edge weights used in the calculation are set to 1 unless the neighborhoods of $i$ and $j$ completely overlap. This is due to $o^W_{ij}$ measuring the proportion of \emph{weight} devoted to common neighbors of $i$ and $j$ rather than the proportion of \emph{neighbors} $i$ and $j$ share. In essence, each common neighbor is counted twice when calculating $o^W_{ij}$; once for its tie to $i$ and once for its tie to $j$. For example, if all edges were set to 1 in Figure \ref{fig:schematic}b, the weighted overlap would be $o^W_{ij} = \sum^{n_{ij}}_{k=1}(w_{ik} + w_{jk}) / (s_i + s_j - 2w_{ij}) = ((1+1)+(1+1)) / (5 + 5 - 2) = 1 / 2$,
rather than 1/3 as calculated above.

In a directed network, each edge has a direction associated with it. This makes it possible for ties to be reciprocated, meaning that there can be an edge pointing from node $i$ to $j$ and another edge pointing from $j$ to $i$. For directed networks, the concept of a ``common neighbor'' is ambiguous due to the directionality associated with ties. We define a common neighbor as a node that enables or mediates a directed path of length two between the source and target nodes, either from $i$ to $j$, from $j$ to $i$, or both. Defining a common neighbor in this manner allows information to flow between $i$ and $j$ via a common neighbor. To illustrate this, let $i$ and $j$ be the two connected individuals of interest, and let $k$ represent a potential common friend. If there is a directed edge from $i$ to $k$ and a directed edge from $k$ to $j$, then there is a path a length two from $i$ to $j$ through $k$, and $k$ is considered a common friend. Using this criterion, we define directed overlap in Eq. \eqref{eq:3}  as the proportion of paths of length two between two connected individuals:
\begin{eqnarray}\label{eq:3}
o^D_{ij} =  \frac{ \sum^{n}_{k = 1} (A_{ik}A_{kj} + A_{jk}A_{ki}) }{\text{min}(k_i^{\text{in}}, k_j^{\text{out}}) + \text{min}(k_j^{\text{in}}, k_i^{\text{out}} ) - 1}.
\end{eqnarray}
Here, $A_{ij}$ is the $(i,j)$ element of the directed adjacency matrix, $k_i^{\text{in}}$ $(k_j^{\text{in}})$ denotes the in-degree of node $i$ $(j)$, $k_i^{\text{out}}$ $(k_j^{\text{out}})$ denotes the out-degree of node $i$ $(j)$, and min$(\cdot,\cdot)$ the minimum of the two arguments. We consider each edge separately, even in the case of unreciprocated edges, and again, the tie between nodes $i$ and $j$ is not included in the calculation. The metric is only defined if $\text{min}(k_i^{\text{in}}, k_j^{\text{out}}) + \text{min}(k_j^{\text{in}}, k_i^{\text{out}}) > 1$. For nodes $i$ and $j$ in Figure \ref{fig:schematic}c, the directed overlap can be calculated as $o^D_{ij} = \sum^{n}_{k = 1} (A_{ik}A_{kj} + A_{jk}A_{ki}) / (\text{min}(k_i^{\text{in}}, k_j^{\text{out}}) + \text{min}(k_j^{\text{in}}, k_i^{\text{out}} ) - 1) = (1+0) / (2 + 2 - 1) = 1 / 3$.

Note that according to our definition of a common neighbor in a directed network, only node $k$ is a common neighbor of $i$ and $j$ since a directed path of length two does not exist from node $i$ to node $j$ (or from node $j$ to node $i$) through node $h$. 

\vspace{-20pt}

\subsection{\er Random Graph Models}

With the extensions of edge overlap defined above, one can easily compute the mean overlap (simple or weighted or directed) across all edges in the network. However, in order to make meaningful comparisons, such as to learn whether the observed value of edge overlap is small or large, or whether it represents a statistically significant deviation from what might be expected to occur at random, one needs to consider suitable null models and derive both the expected value and the variance of overlap under these null models. The \er random graph model, often denoted $G(n,p)$, is the simplest model for generating random graphs \cite{Erdos}. In this model, graphs are created by considering each of the ${n \choose 2}$ distinct pairs of $n$ nodes and any such pair of nodes is connected with probability $p$ independently of all other dyads (node pairs). The random process can therefore be thought of as a series of Bernoulli trials or coin flips. Suppose we have a coin that lands on heads with probability $p$. If the coin flip results in heads, we connect the two nodes, otherwise we leave them unconnected. Note that since the probability of creating an edge is fixed, but the number of edges is not, the number of edges present in any graph realization is a random variable.

\par The weighted random graph (WRG) is the weighted counterpart of the canonical \er random graph \cite{Diego}. In this case, a network of $n$ nodes is generated by selecting each pair of nodes in turn and carrying out a series of independent Bernoulli trials for each pair with success probability $p$. This process is continued until the first failure is encountered, and every success preceding the failure results in the addition of a unit weight to the tie. Note that if the first Bernoulli trial is a failure, the two nodes will not be connected. We can again relate the graph construction to the tossing of a coin. If the coin lands on heads with probability $p$, the weight associated with an edge is given by the number of heads flipped until getting the first tails, and therefore tie weights are distributed according to the geometric distribution. This process is repeated independently for every distinct pair of nodes in the network. 

\par The directed random graph is the directed version of the \er random graph, and it is generated in a very similar manner as its canonical counterpart. For two nodes $i$ and $j$ in a network of $n$ nodes, an edge pointing from $i$ to $j$ is created with probability $p$ and, likewise, an edge pointing from $j$ to $i$ is also connected independently with probability $p$ \cite{Erdos, Erdos2, Ballobas}. In this case, in the coin analog of the model, we flip a coin twice for each ordered pair (rather than unordered pair) of nodes, one flip for each direction. This process is repeated independently for every pair of nodes in the network. We note in passing that it appears that one could naturally generalize the \er graph to a weighted and directed network by carrying out the process described for WRG but doing so separately for each ordered pair of nodes.


\subsection{\er Overlap}
In order to perform inference about edge overlap, i.e., to compare point estimates of overlap across networks, we need to know the mean and variance of each version of overlap under the null model in question. To fix our notation, we will let uppercase letters stand for random variables: $K_i$ denotes the degree of node $i$, $N_{ij}$ the number of common neighbors of nodes $i$ and $j$, $S_i$ the strength of node $i$, $W_{ij}$ the weight of the edge connecting nodes $i$ and $j$, $K^{in}_i$ the in-degree of node $i$, $K^{out}_i$ the out-degree of node $i$, and $A_{ij}$ the adjacency matrix element, where a nonzero (positive) value represents the existence of an edge between nodes $i$ and $j$ (binary in the case of unweighted graphs).

For the \er random graph, a given node is connected to each of the remaining $n-1$ nodes with probability $p$, and its resulting degree can thus be viewed as a sum of independent Bernoulli trials. Therefore, as is well known, $K_i \sim$ binomial$(n-1, p)$, which can be approximated by a Poisson$(np)$ distribution for large $n$. For any pair of (connected) nodes, the probability of both nodes being connected to the same neighboring node, meaning that they have a common neighbor, is $p^2$ as each edge occurs independently of any other. Moreover, the total number of possible common friends two nodes can have is $n - 2$. Thus, $N_{ij} \sim$ binomial$(n-2, p^2)$, which can similarly be approximated by a Poisson$(np^2)$ random variable for large $n$. With these definitions, the numerator of edge overlap is a Poisson random variable, and the denominator is the difference of two Poisson random variables, known as a Skellam random variable \cite{Skellam}. In this case, the denominator is a Skellam$(2np - 2 - np^2)$ random variable. We can now view overlap as a random variable as in Eqn. \eqref{eq:unweightedrv}.

\begin{equation}\label{eq:unweightedrv}
O_{ij} = \frac{N_{ij}}{(K_i -1) + (K_j -1) - N_{ij}}
\end{equation}

Edge overlap is a ratio of two dependent random variables since the maximum number of possible common friends is bounded by the min$(K_i, K_j)$. This dependency increases the difficulty of deriving exact expressions for the mean and variance of overlap. However, despite this dependence, we can approximate both the mean and variance in two different ways. The first approach observes the weakness of the dependence between the numerator and denominator and simply ignores it, defining the ratio as a function of two independent random variables. Approximations for the mean and variance of the ratio are then derived using Taylor expansions of the function about the means of the random variables \cite{Kendall, Johnson}. This results in Eqs. (5) and (6) (for details, see Appendix A.1.).

\begin{eqnarray}
\mathbb{E}[O_{ij}] &=& \frac{p}{2-p} \\[15pt]
\text{Var}(O_{ij}) &=&  \frac{np^2}{(2np - 2 - np^2)^2} + \frac{n^2p^4(2np - 2 + np^2)}{(2np - 2 - np^2)^4}.
\end{eqnarray}

Our second approach incorporates results from \cite{Oxford}, where the local clustering coefficient for an \er random graph is also written as a ratio of two dependent random variables with the intention of deriving its distribution. The dependency is eliminated by replacing the random variable in the denominator with its expectation, and this approximation turns the denominator into a constant. Thus, the distribution of the clustering coefficient is approximated by a scaled version of the random variable in the numerator. It is subsequently shown that this is a good approximation for the actual distribution. We adopt this approach here as our second approach, and approximate the distribution of edge overlap by replacing the denominator with its expectation. We then derive the mean and variance of $O_{ij}$ using the distributional properties of the numerator. This results in the expressions in Eqs. (7) and (8) (for details, see Appendix B.1.): 
\begin{eqnarray}
\mathbb{E}[O_{ij}] &=& \frac{p}{2-p} \\[15pt]
\text{Var}(O_{ij}) &=& \frac{np^2}{(2np - 2 - np^2)^2}.
\end{eqnarray}
Note that the expressions for the mean using the two approaches in Eqs. (5) and (7) are equivalent, but the expressions for the variance in Eqs. (6) and (8) differ, with the expression for Eq. (8) corresponding to the first term of Eq. (6).

We use the same two approaches for the weighted and directed cases. For the weighted \er random graph (WRG), we first define the distributions of $W_{ij}$ and $S_i$. Given how WRGs are constructed (see above), the tie weights follow a geometric distribution, such that if an edge is placed between a pair of nodes with probability $p$, tie weight distribution will be $W_{ij} \sim$ geometric$(1-p)$. It then follows that node strength $S_i$ is a sum of geometric random variables, i.e., is the sum of the weights of the ties that are adjacent to the given node, leading to $S_i \sim$ negative binomial$(n-1, 1-p)$ \cite{Diego}. (Note that we only consider integer weights in our derivations.)

For the first approach, the numerator can be written as $\sum^{N_{ij}}_{k=0} (W_{ik} + W_{jk})$, where $N_{ij}$ is again the number of common neighbors of nodes $i$ and $j$, and is distributed as in the unweighted \er random graph. Thus, the numerator is a sum of geometric random variables, where the number of summed variables is itself a random variable. Moreover, we must have $W_{ik} > 0$ and $W_{jk} >0 $ since a common neighbor of two nodes can only exist if both nodes are attached to the node in question (the common neighbor). To address this constraint, each of the random variables must first be transformed into zero-truncated geometric random variables, and their mean and variance altered correspondingly. We can now write weighted overlap as a random variable as in Eqn. \eqref{eq:weightedrv}.
\begin{eqnarray}\label{eq:weightedrv}
O^W_{ij} = \frac{\sum^{N_{ij}}_{k=1}(W_{ik} + W_{jk})}{S_i + S_j - 2W_{ij}}.
\end{eqnarray}
Hierarchical models can now be used to find the mean and variance of the numerator, and these results combined with the mean and variance values of the denominator can be used to derive the expressions in Eqs. (10) and (11) (see Appendix A.2 for details): 

\begin{eqnarray}
\mathbb{E}[O^W_{ij}] &=& p\\[15pt]
\text{Var}(O^W_{ij}) &=& \frac{p+1}{n} - \frac{p(1-p^2)(1 - p(p^2-3p+3))}{(np-1)}.
\end{eqnarray}

The second approach again replaces the denominator with its expectation. The mean and variance derivations are then straightforward and result in the expressions in Eqs. (12) and (13). Again, the expressions for the mean are equivalent for both approaches, and the variance expressions are similar but not identical (See Appendix B.2 for the details):

\begin{eqnarray}
\mathbb{E}[O^W_{ij}] &=& p\\[15pt]
\text{Var}(O^W_{ij}) &=& \frac{np^2(p+2)}{2(np-1)^2}.
\end{eqnarray}

The derivations for the directed \er random graph are more complicated and do not appear to have a closed form due to the minimum expressions in the denominator. Focusing on the numerator, each of the $A_{ik}A_{kj}$ and $A_{jk}A_{ki}$ terms is equal to one if and only if both adjacency matrix values are equal to 1, which happens with probability $p^2$ since each edge is independent. Thus, each of the terms is a Bernoulli$(p^2)$ random variable, and the numerator consists of a sum of 2$n$ independent Bernoulli random variables, meaning it is a binomial$(2n, p^2)$ random variable, which we will again approximate with a Poisson$(2np^2)$ random variable. The denominator includes the minimum of two identically distributed random variables $K^{in}_i$ and $K^{out}_i$. Due to the definition given in Section 3.1, the in and out degrees of nodes $i$ and $j$ cannot equal 0, making them zero-truncated binomial$(n-1, p)$ random variables, which will also be approximated as zero-truncated Poisson$(np)$ random variables since $n$ is assumed to be large. We can now write directed overlap as a random variable as in Eqn. \eqref{eq:directedrv}.

\begin{eqnarray}\label{eq:directedrv}
O^D_{ij} =  \frac{ \sum^{n}_{k = 1} (A_{ik}A_{kj} + A_{jk}A_{ki}) }{\text{min}(K_i^{\text{in}}, K_j^{\text{out}}) + \text{min}(K_j^{\text{in}}, K_i^{\text{out}} ) - 1}.
\end{eqnarray}

The mean and variance of the denominator can now be calculated and used to derive the expressions in Eqs. (15) and (16) \cite{Kendall} (for details, see Section Appendix A.3.): 

\begin{eqnarray}
\mathbb{E}[O^D_{ij}] &=& \frac{np^2}{e^{-2np}\sum^{(n-1)}_{k=1} \left[\sum^{(n-1)}_{j=k} \frac{(np)^j}{j!} \right]^2 - 0.5} \\ [15pt]
\text{Var}(O^D_{ij}) &=& \frac{2n^2p^4}{\left(2e^{-2np}\sum^{(n-1)}_{k=1} \left[\sum^{(n-1)}_{j=k} \frac{(np)^j}{j!} \right]^2 - 1\right)^2} \\ [15pt]
&+& \frac{\frac{32n^3p^5e^{np}}{e^{np} - 1}\left[1 - \frac{np}{e^{np} -1}\right] }{\left(2e^{-2np}\sum^{(n-1)}_{k=1}\left[\sum^{(n-1)}_{j=k} \frac{(np)^j}{j!}\right]^2 - 1\right)^2}. \nonumber
\end{eqnarray}

The second approach again replaces the denominator with its expectation, and the mean and variance derivations result in the expressions in Eqs. (17) and (18) (see Appendix B.3 for details). Again, the expressions for the mean are equivalent for both approaches, but the expression for the variance using the second approach in Eq.~18 is equivalent to the first term of the variance resulting from the first approach in Eq.~16.

\begin{eqnarray}
\mathbb{E}[O^D_{ij}] &=& \frac{np^2}{e^{-2np}\sum^{(n-1)}_{k=1}\left[\sum^{(n-1)}_{j=k} \frac{(np)^j}{j!} \right]^2 - 0.5} \\[15pt]
\text{Var}(O^D_{ij}) &=& \frac{2n^2p^4}{(2e^{-2np}\sum^{(n-1)}_{k=1}\left[\sum^{(n-1)}_{j=k} \frac{(np)^j}{j!} \right]^2 - 1)^2}
\end{eqnarray}


\section{Simulation Studies}
We conducted simulation studies to evaluate the accuracy of the proposed mean and variance expressions for each version of \er edge overlap. We simulated 5,000 realizations of networks with $n =1,000$ nodes in each for various values of $p \in (0,1)$. The mean and variance of edge overlap was calculated for each network realization, and those values were subsequently averaged over all simulations. We considered values of $p > 1/n$, resulting in average degree $np > 1$, and ensuring (asymptotically) that the graphs have non-vanishing giant components. Note that $np$ is the average in-degree as well as the average out-degree for directed networks.

\begin{figure}
     \begin{center}
        \subfigure[]{%
            \includegraphics[width=0.3\textwidth]{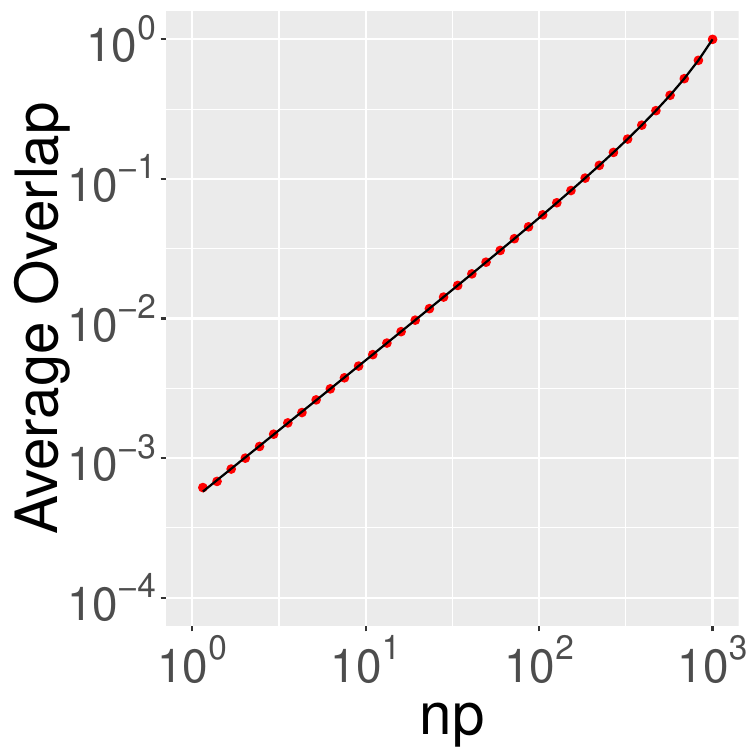}
        }%
        \subfigure[]{%
           \includegraphics[width=0.3\textwidth]{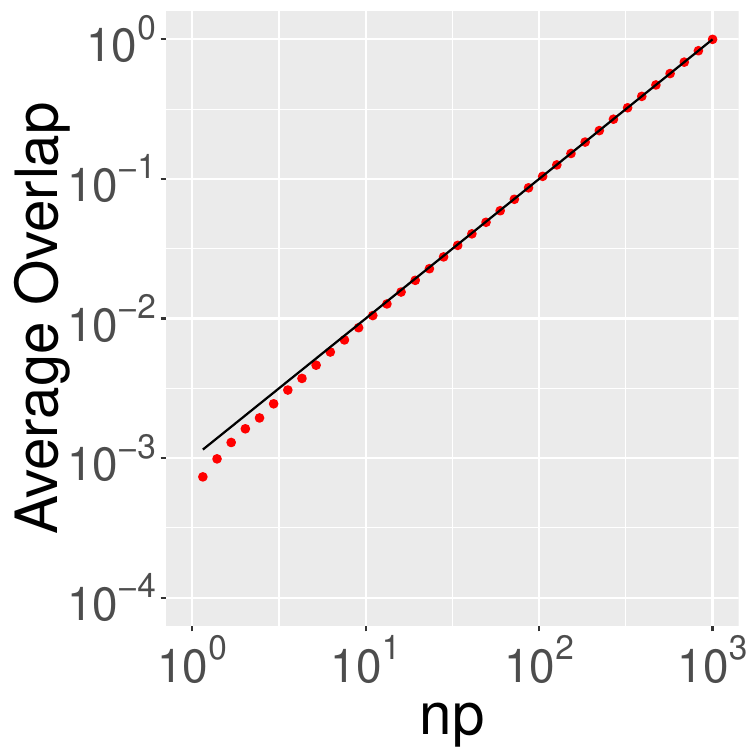}
         }%
         \subfigure[]{%
           \includegraphics[width=0.3\textwidth]{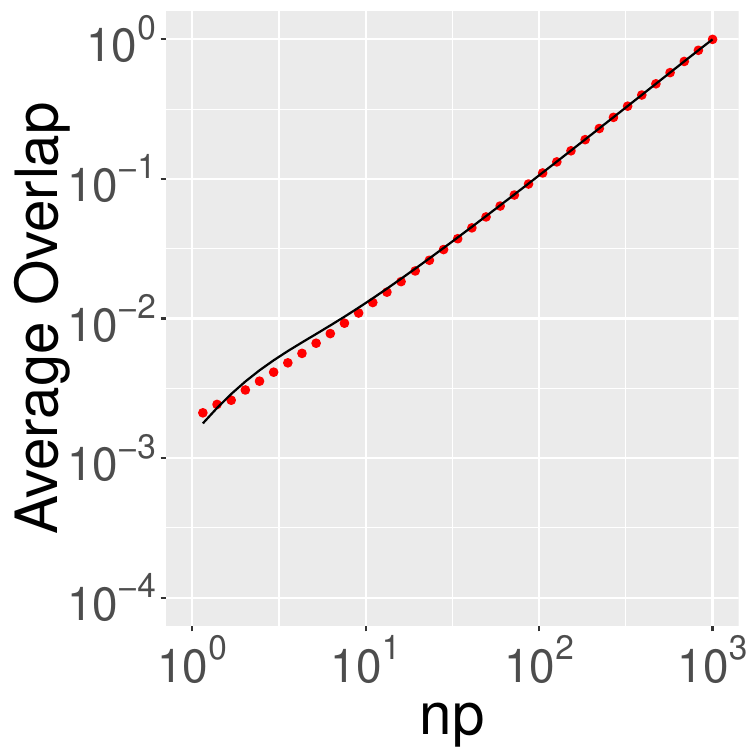}
        }\\ 
        \subfigure[]{%
            \includegraphics[width=0.3\textwidth]{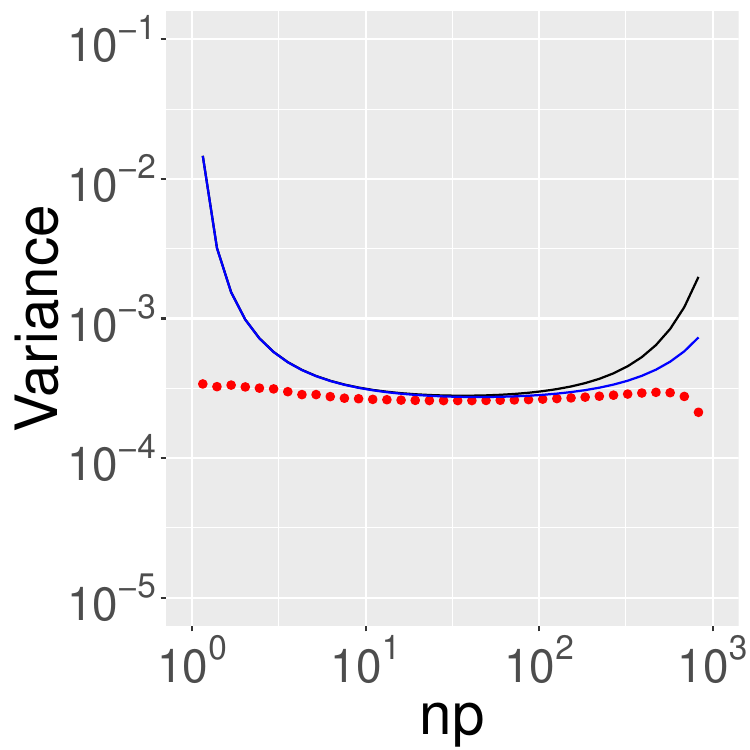}
        }%
        \subfigure[]{%
            \includegraphics[width=0.3\textwidth]{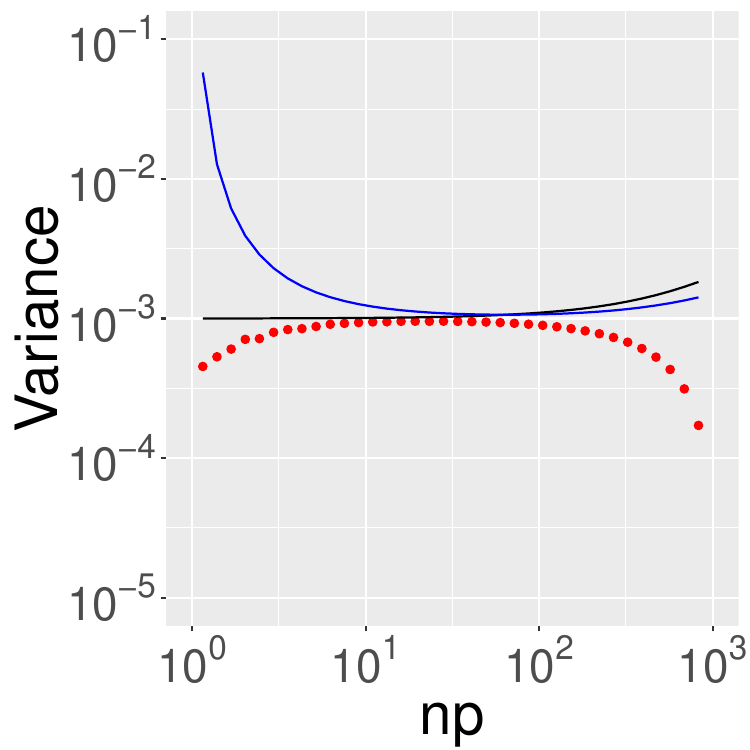}
        }%
         \subfigure[]{%
            \includegraphics[width=0.3\textwidth]{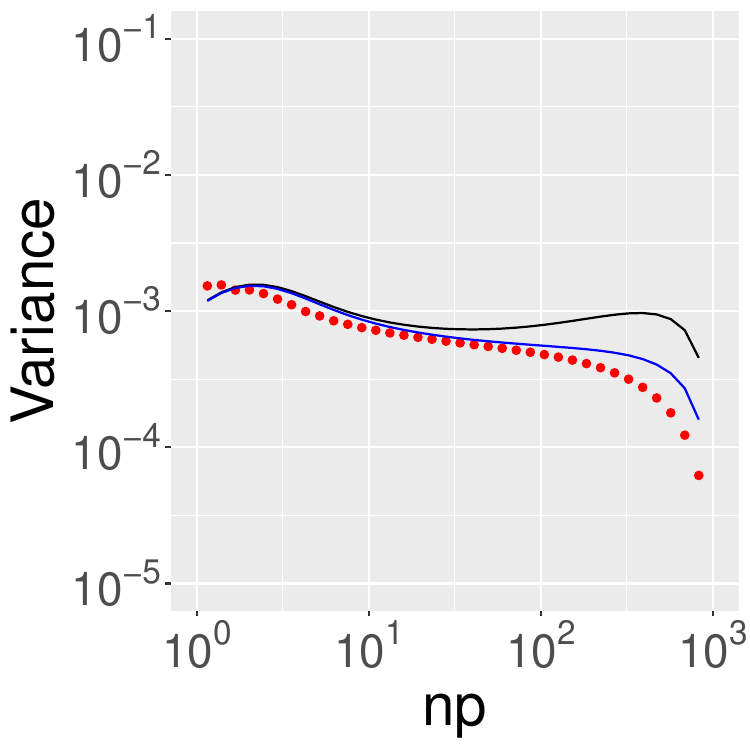}
        }%
    \end{center}
    \captionsetup{width=\textwidth}
    \caption{%
Simulation results for the mean (top row) and variance (bottom row) of each type of \er overlap. The first column corresponds to the unweighted \er overlap, the second column to the weighted \er overlap, and the third column to the directed \er overlap. The top row plots (a), (b) and (c) display the average overlap on the $y$-axis and average degree ($np$) on the $x$-axis. The red dots represent values from the simulations, and the black lines represent the theoretical results using the first approach and the blue lines represent the theoretical outcome using the second approach. Note that the blue lines are completely covered by the black lines since the values for average overlap are the same for both approaches. The bottom row plots (d), (e) and (f) show the variance of edge overlap on the $y$-axis and average degree ($np$) on the $x$-axis using the same colors as the other panels.}%
   \label{fig:sims}
\end{figure}

Figure \ref{fig:sims} displays the simulation results and accuracy of our approximations. The top row contains the results for the mean unweighted overlap (Figure \ref{fig:sims}a), mean weighted overlap (Figure \ref{fig:sims}b), and mean directed overlap (Figure \ref{fig:sims}c). In each plot, the red dots represent the simulated results, black lines represent the theoretical results using the first approach, and blue lines the theoretical results using the second approach. Note that each expression for average overlap is equivalent for the two approaches, making only the black lines visible. The bottom row of panels shows the results for the variance of unweighted overlap (Figure \ref{fig:sims}d), weighted overlap (Figure \ref{fig:sims}e), and directed overlap (Figure \ref{fig:sims}f) using the same markers and line styles as above.

For each version of overlap, our theoretical approximations of the mean closely match the simulations, with the unweighted case being the best overall fit for all values of $np$. The approximations of the variance are not as accurate, and the accuracy depends on the value of $np$. In the unweighted case (Figure \ref{fig:sims}d), both theoretical approaches match the simulated values for average degree $np \geq 10$ until about $np = 100$. The first approximation then deviates from the simulated values, followed by the second approach deviating when $np \approx 300$. In the weighted case (Figure \ref{fig:sims}e), the first approximation is more accurate than the second until average degree is about 30. The approaches are then equally accurate until the average degree reaches 170, and after this point the second approximation matches the simulated results more closely. In the directed case (Figure \ref{fig:sims}f) the two approximations are equivalent and closely match the unique shape of the theoretical values. Only for very small values of $np$ do the approximations slightly underestimate the variance. After that point, both approaches are very similar and approach two is more accurate for $np > 10$. Furthermore, in all cases, both approximations systematically overestimate variability. We stress that this overestimation leads to inflated standard errors and thus to conservative hypothesis tests, which is preferable over the opposite situation, i.e., having deflated standard errors and anti-conservative tests. 

Overall, our approximations are accurate for values of average degree seen in many empirical networks; namely $5 \geq np \leq 300$. The approximations from one or both approaches are only slightly higher for most of the values of average degree, and deviate substantially only when $np$ = 1, which marks the location of the percolation transition, or $np >$ 1000, which introduces complicated dependencies that are not captured accurately by our approximations. 

To investigate the performance of our derivations when used for hypothesis testing we calculated the coverage probability of 95\% confidence intervals constructed for each type of overlap and both of the approximation approaches. We first created an \er network of 500 nodes and average degree ($np$) equal to 5, and calculated edge overlap for each edge in the network. Using our first approach mean and variance equations we determined the 95\% confidence interval upper and lower bounds for the network and computed the proportion of edge overlap values that fell within this interval. We repeated this process 100 times each for $np$ = 5, 10, 25 and 100, with the final coverage probability being the average of the probabilities across the 100 networks. This process was then replicated using the derivations from approach 2. Finally, all steps were repeated for networks with 1,000 nodes and the weighted and directed versions of the \er random graph. The resulting coverage probabilities are shown in Table 1.

Almost all coverage probabilities are greater than or equal to 95\%, suggesting our derivations are accurate and would produce conservative hypothesis tests. This is in agreement with the simulations for mean and variance shown above, especially for small and large $n$ where our approximations are less precise. Of the 48 coverage probabilities presented in Table 1, only 3 were less than 0.95, resulting in an anti-conservative test. All three result when $np = 5$ for directed networks. This is expected since our variance approximations for directed \er networks is smaller than it should be for small values of $n$.

\setlength{\tabcolsep}{10pt}
\begin{table}[t]
\centering
\begin{tabular}{cllcccc}
 & & & \multicolumn{4}{c}{$np$} \\
\cline{4-7}
Derivation  & Overlap &  & &  & &  \\
Approach & Type & $n$ & 5 &  10  & 25 & 100 \\
 \hline
& Original ($O_{ij}$)            & 500     &  0.9528  & 0.9539 & 0.9625 & 0.9743	  \\
               		   &              & 1,000  & 0.9748  & 0.9756 & 0.9598 & 0.9647	  \\ \\
1 & Weighted ($O^W_{ij}$)  & 500     & 0.9522  & 0.9525	& 0.9655 & 0.9840	  \\
                		   &              & 1,000  &  0.9742 & 0.9560	& 0.9578 & 0.9713	  \\ \\  
& Directed  ($O^D_{ij}$)      & 500     & 0.9375   & 0.9588 & 0.9816 & 0.9970  \\
               &                           & 1,000  & 0.9524 & 0.9559 & 0.9696 & 0.9876  \\ 
\hline

& Original ($O_{ij}$)            & 500      & 0.9537 & 0.9528 & 0.9609 & 0.9640	  \\
                                      &    & 1,000   & 0.9760 & 0.9577	& 0.9555 & 0.9607	  \\ \\
2 & Weighted ($O^W_{ij}$)   & 500      & 0.9546 & 0.9612	& 0.9675 & 0.9802	  \\
                                       &   & 1,000   &  0.9747& 0.9533	& 0.9613 & 0.9693	  \\ \\
& Directed ($O^D_{ij}$)         & 500      & 0.9258 & 0.9537 & 0.9727 & 0.9744  \\
                                       &   & 1,000   &0.9484  & 0.9507 & 0.9640  & 0.9644 \\ 
\end{tabular}
\caption{Results of coverage probability simulations for 95\% nominal confidence intervals. Each value is the resulting average coverage probability from 100 random networks. A value of 0.95 would be ideal; here almost all probabilities greater than or equal to 0.95, producing conservative hypothesis tests. Three values are less than 0.95 and would result in anti-conservative hypothesis tests, showing that our approximations are less precise for small values of $n$.}
\label{table: coverage}
\end{table}

\section{Data Analysis}

As an application of our derivations to the analysis of empirical social networks, we used social network data collected in 2006 from 75 villages housed in 5 districts in rural southern Karnataka, India \cite{Banerjee}. The data were collected as part of a study that examined how participation in a microfinance program diffuses through social networks. First, a baseline survey was conducted in all 75 villages. The survey consisted of a village questionnaire, a full census that collected data on all households in the villages, and a detailed follow-up survey fielded to a subsample of individuals. The village questionnaire collected data on village leadership, the presence of pre-existing non-governmental organizations (NGOs) and savings self-help groups and various geographical features of the area. The household census gathered demographic information, GPS coordinates of each household and data on a variety of amenities for every household in each village (roof type, latrine type, and access to electric power). The individual surveys were administered to a random sample of villagers in each village and were stratified by religion and geographic sub-location. Over half of the households in each stratification were sampled, yielding a sample of about 46$\%$ of all households per village. The individual questionnaire asked for information including age, sub-caste, education, language, native home, and occupation of the person. Additionally, the survey included social network data along 12 dimensions: friends or relatives who visit the respondent's home, friends or relatives the respondent visits, any kin in the village, nonrelatives with whom the respondent socializes, those who the respondent receives medical advice from, who the respondent goes to pray with, from whom the respondent would borrow money, to whom the respondent would lend money, from whom the respondent would borrow or to whom the respondent would lend material goods, from whom the respondent gets advice, and to whom the respondent gives advice. 

The median pairwise distance between villages was 46km and the number of cross-village ties was minimal, allowing the villages to be regarded as independent networks. Each village contained anywhere from 354 to 1,775 residents, with a total population of 69,441 people in the 75 villages combined. The number of edges across all social networks totaled 2,361,745 which included 480 self-loops and 6,402 isolated dyads. The average degree was 6.79 (standard deviation of 4.03), and the average number of connected components was 17.99 per village.

We first calculated the average unweighted overlap for each type of social relationship (labeled 1-12, see Table \ref{table: relations}) for each village by treating all ties as undirected and by removing all self-loops since they do not contribute to edge overlap (Figure \ref{fig:full_raw}). Then we standardized each average overlap by subtracting the expected mean and dividing by the standard deviation under the null; the results from the unweighted \er overlap derivations using the first approach discussed above (Figure \ref{fig:full_stand}). 

We next collapsed the twelve unweighted networks into one weighted network. Specifically, the weight of a tie between two individuals corresponds to the number of types of social relationships they are engaged in with each other. For example, if individual $i$ borrows money from, gives advice to and goes to temple with individual $j$, the weight of the (undirected) tie between $i$ and $j$ would be equal to 3. Figure \ref{fig:full_w} shows the distributions of raw and standardized weighted overlap for all edges. 

\begin{table}[htb]
\centering
\begin{tabular}{ccl}
  Label & \hspace{0.75cm}  & Type of social interaction \\
\hline
  1 &&  The respondent borrows money from this individual\\
  2 && The respondent gives advice to this individual \\
  3 && The respondent helps this individual make a decision \\
 4 && The respondent borrows kerosene or rice from this individual\\
 5 && The respondent lends kerosene or rice to this individual\\
 6 && The respondent lends money to this individual\\
 7 && The respondent obtains medical advice from this individual\\
 8 && The respondent engages socially with this individual \\
 9 && The respondent is related to this individual\\
 10 && The respondent goes to temple with this individual\\
 11 && The respondent has visited this individual's home\\
 12 && The respondent has been invited to this individual's home\\
\end{tabular}
\caption{The types of social interactions recorded for individuals in each village. }
\label{table: relations}
\end{table}

\section{Results}
\begin{figure}
\vspace{10pt}
\centering
\includegraphics[scale = 0.4]{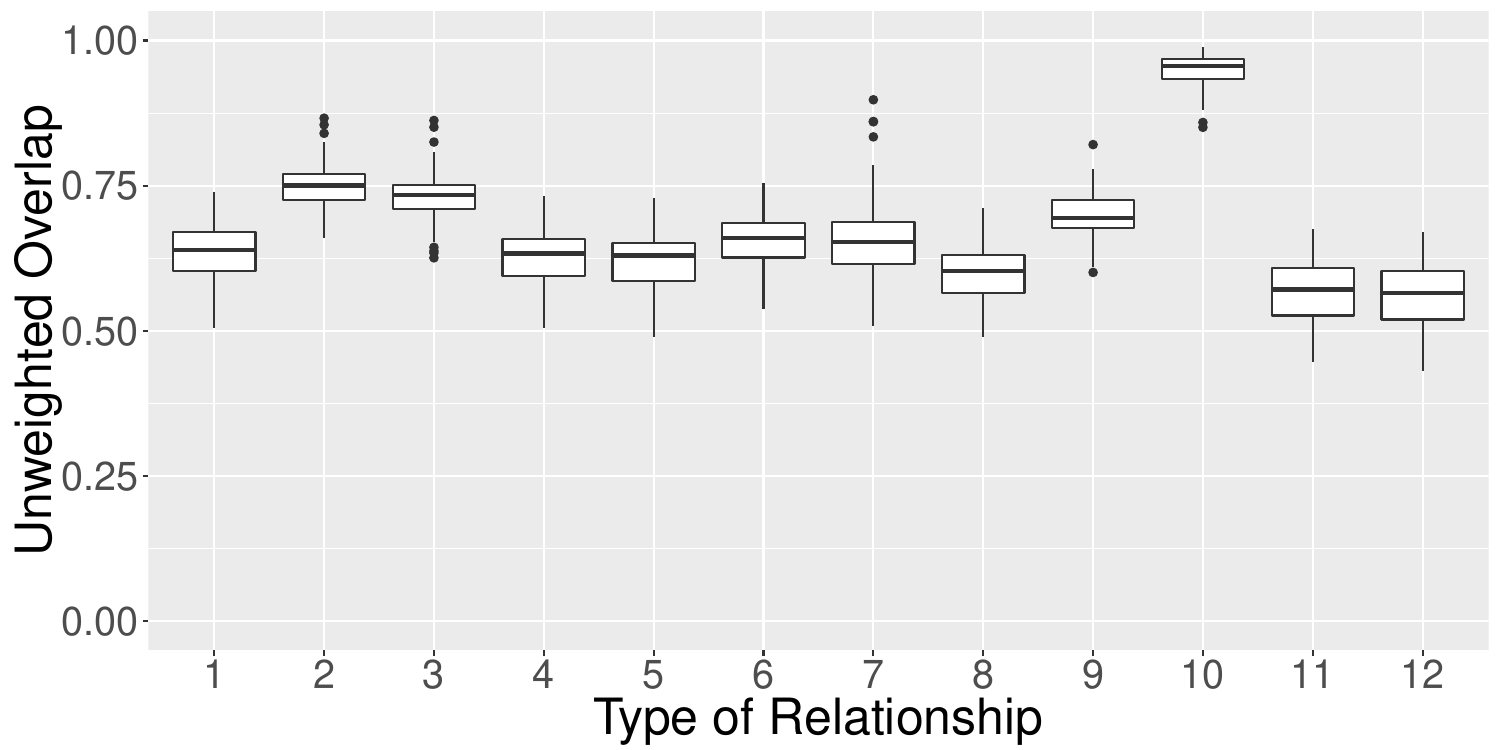}
\captionsetup{width=\textwidth}
\caption{Distribution of average unweighted overlap for each type of social interaction. The average overlap was calculated for each type of interaction for each of the 75 villages. The y-axis represents the proportion of average edge overlap and the x-axis represents the type of social interaction. See Table \ref{table: relations} above for full descriptions interaction types. }
\label{fig:full_raw}
\end{figure}

\begin{figure}
\centering
\includegraphics[scale = 0.4]{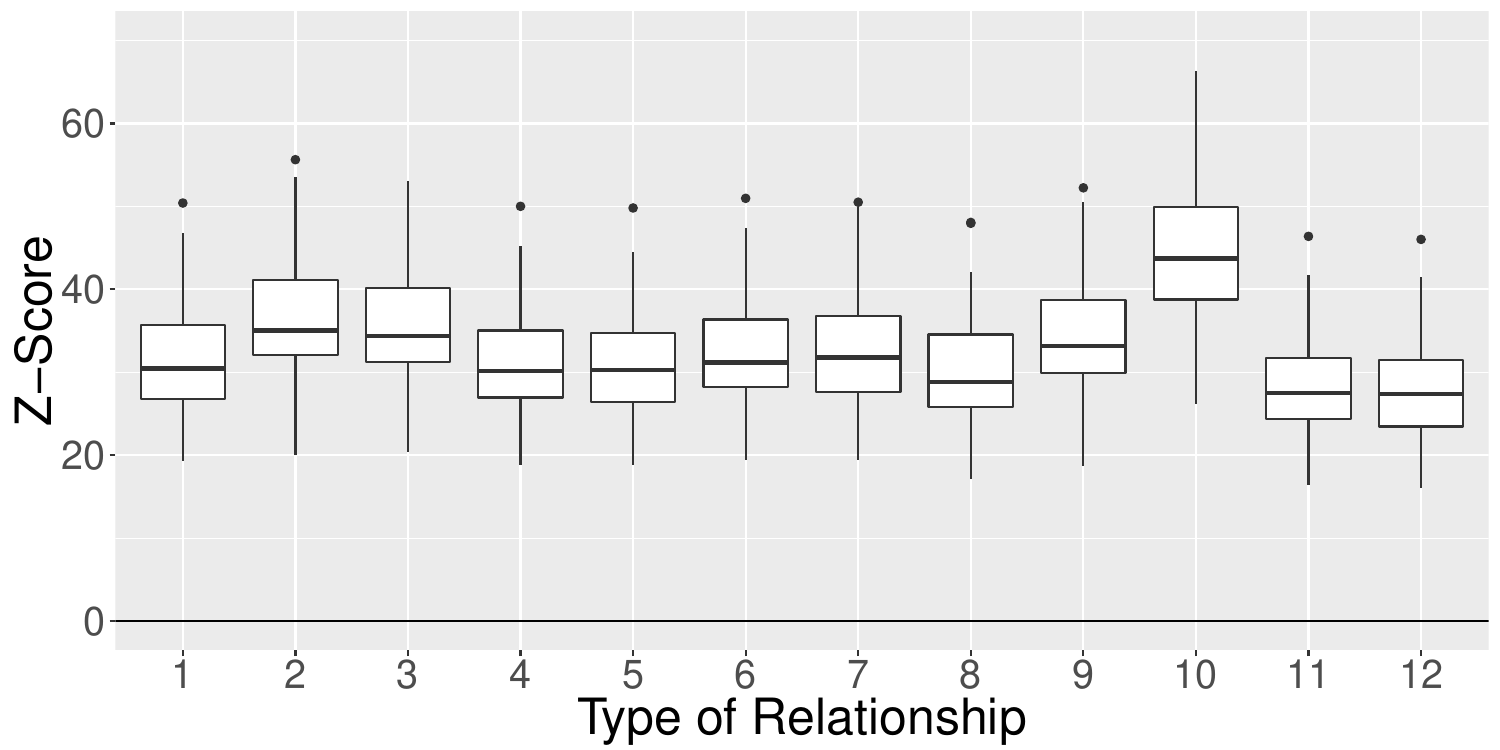}
\captionsetup{width=\textwidth}
\caption{Distribution of standardized unweighted overlap for each village for each type of social interaction. Using the approximations from Approach 1, each standardized value was calculated by first subtracting the expected mean overlap under the null from the observed average overlap (the values in Figure \ref{fig:full_raw}), and then dividing that value by the expected standard deviation under the null. The y-axis represents the standardized value, also known as the Z-score, and the x-axis represents the type of social interaction. See Table \ref{table: relations} above for full descriptions interaction types.}
\label{fig:full_stand}
\end{figure}

Figure \ref{fig:full_raw} illustrates the average raw unweighted overlap for each type of social interaction for each village. Each distribution is fairly normally distributed with the exception of interaction types 2, 7 and 10. Each distribution also showcases minimal variance and medians above 0.5. It is also clear that the values of average overlap for social interaction type 10 are very large and could indicate the importance of attending temple among these villages. Figure \ref{fig:full_stand} shows the distributions of the standardized unweighted overlaps. Clearly, every value of average unweighted overlap is significantly larger than expected under the null of a random network; the minimum values for each type of interaction never fall below 10 standard deviations from the mean, and the maximum value is greater than 60 standard deviations from the mean.  Again, the values from interaction type 10 are among the largest values, suggesting that villagers who attend temple together have a significantly higher proportion of mutual friends compared to other types of interaction and the null model. Values significantly higher than expected under the null are not unusual since social networks are known to have a larger amount of clustering compared to random graphs due to different social mechanisms that drive the formation of clustered ties. Additionally, the \er random graph model is the simplest null model with no clustering by design, and is easily rejected when analyzing empirical social networks.

\par
\begin{figure}
\begin{minipage}[b]{.4\textwidth}
\centering
\includegraphics[width=\textwidth]{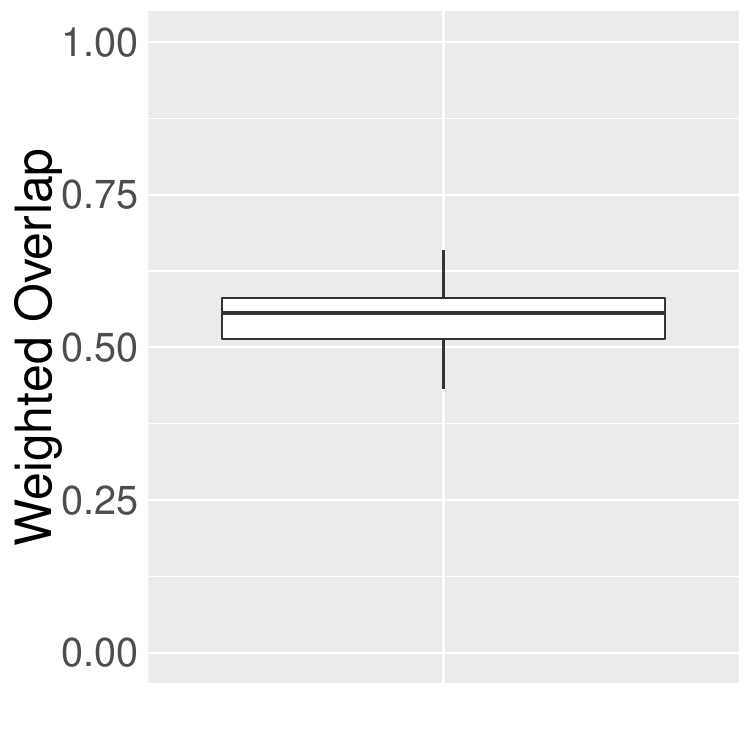}
\caption*{(a)}
\label{fig:full_raw_w}
\end{minipage}
\hspace{2cm}
\begin{minipage}[b]{.4\textwidth}
\centering
\includegraphics[width=\textwidth]{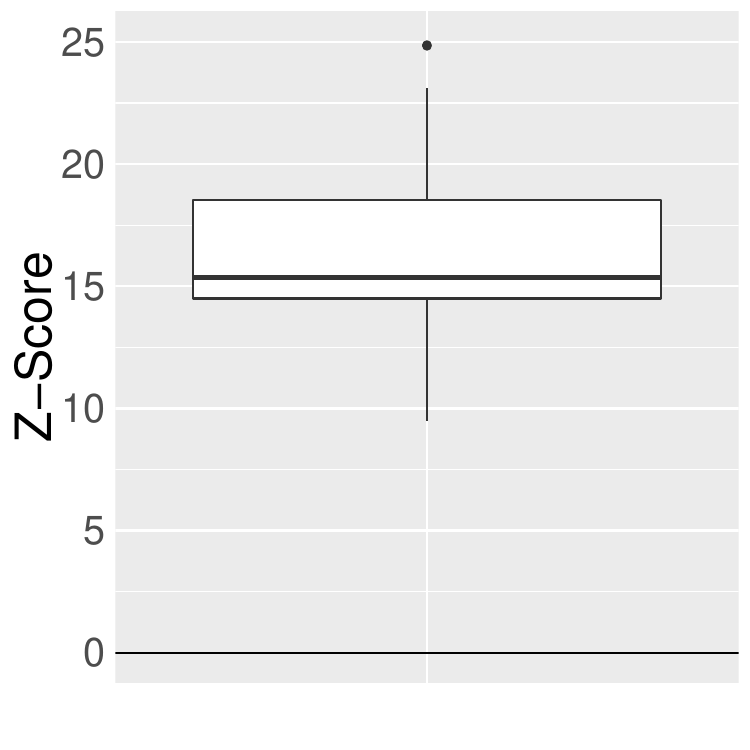}
\caption*{(b)}
\label{fig:full_stand_w}
\end{minipage}
\captionsetup{width=\textwidth}
\caption{Distribution of average weighted overlap (a) and standardized weighted overlap (b) for all villages. We created a weighted network for each village by collapsing the twelve unweighted interaction networks. The weight of an edge between two individuals corresponds to the number of social interactions they engage in. The y-axis in (a) represents the proportion of average weighted edge overlap, and the y-axis in (b) represents the standardized value, also known as the Z-score. Using the approximations from Approach 1, each standardized value was calculated by first subtracting the expected mean overlap under the null from the observed average overlap (the values in (a)), and then dividing that value by the expected standard deviation under the null.}
\label{fig:full_w}
\end{figure}

 The distribution of average weighted overlap (Figure \ref{fig:full_w}a) is normally distributed with a mean of 0.548 and standard deviation of 0.046. Each village's average weighted overlap is significantly different from what is expected under each corresponding null value (Figure \ref{fig:full_w}b). This is expected given the values in Figure \ref{fig:full_stand} for each type of social interaction are also significantly higher than expected, and that humans do not typically create friendships randomly.

\section{Conclusions and Discussion}

In this paper, we introduced extensions of edge overlap for weighted and directed networks. We also used the classic \er random graph and its weighted and directed counterparts to define a null model and derive approximations for the expected mean and variance of edge overlap for each type of graph. Edge overlap can be standardized using these approximations, which allows its principled comparison across networks of different size. We used these approximations in a data analysis of the social networks of 75 villages in rural India. We found that overall, the average proportion of overlap was much higher than expected under the null for each type of social interaction, especially when the social activity was going to temple together. While our work generalizes an important microscopic network metric, making it more broadly applicable, there are limitations to our work. 

We define extensions of edge overlap to weighted and directed networks that are directly motivated by Granovetter's canonical work on the role of weak ties for social networks. However, other definitions are certainly possible and we hope our work provides a foundation for other researchers to modify as needed. While our estimates of the mean are quite accurate, our variance estimates break down at the boundaries of average degree. That they break down for very sparse or very dense networks is perhaps not surprising; many statistical methods breakdown at the boundaries of the parameter space. A classic example of this is maximum likelihood estimation. These boundaries are typically outside of the average degree seen in social networks, but derivation of variance estimators that work better in these more extreme settings presents an interesting avenue for future work. Another limitation with our mean and variance approximations is the fact that we ignore the correlations that are present among the random variables in the overlap expressions. While our approximations are reasonably precise in most cases, they could be improved if the correlation were taken into account in the approximations, especially for values of average degree less than 5 and greater than 300. Finally, the \er random graph model is a simple and somewhat naive null model in the context of social networks. This model does not preserve the degree distribution and is relatively easy to reject. An alternative would be to derive these expressions for the configuration model, which does preserve the degree distribution. Deriving the mean and in particular the variance under the configuration null model would be considerably more difficult, but we hope our work provides inspiration and direction for future work in this area.


\section*{Conflicts of Interest}
The authors have nothing to disclose.

\clearpage
\bibliographystyle{abbrv}
\bibliography{p1bib}{}

\begin{thebibliography}{10}

\bibitem{Banerjee}
A.~Banerjee, A.~Chandrasekhar, E.~Duflo, and M.~Jackson.
\newblock The diffusion of microfinance.
\newblock {\em Science}, 341, 2013.

\bibitem{Bianconi}
G.~Bianconi, R.~Darst, J.~Iacovacci, and S.~Fortunato.
\newblock Triadic closure as a basic generating mechanism of communities in
  complex networks.
\newblock {\em Physics Review}, 90, 2014.

\bibitem{Ballobas}
B.~Bollob\'{a}s.
\newblock {\em Random Graphs}.
\newblock Academic Press, 1985.

\bibitem{Centola}
D.~Centola.
\newblock {\em How Behavior Spreads: The Science of Complex Contagions}.
\newblock Princeton University Press, 2018.

\bibitem{Chou}
A.-A. Choumane, A. and Harkous.
\newblock Core expansion: a new community detection algorithm based on
  neighborhood overlap, 2020.

\bibitem{Chris}
N.~A. Christakis and J.~H. Fowler.
\newblock The spread of obesity in a large social network over 32 years.
\newblock {\em N. Engl. J. Med.}, 357:370--379, 2007.

\bibitem{Chris2}
N.~A. Christakis and J.~H. Fowler.
\newblock The collective dynamics of smoking in a large social network.
\newblock {\em N. Engl. J. Med.}, 358:2249--2258, 2008.

\bibitem{Johnson}
R.~Elandt-Johnson and N.~Johnson.
\newblock {\em Survival Models and Data Analysis}.
\newblock John Wiley \& Sons, 1998.

\bibitem{Erdos}
P.~Erd\H{o}s and A.~R\'{e}nyi.
\newblock On random graphs i.
\newblock {\em Publicationes Mathematicae}, 6:290--297, 1959.

\bibitem{Erdos2}
P.~Erd\H{o}s and A.~R\'{e}nyi.
\newblock On the evolution of random graphs.
\newblock {\em Publ. Math. Inst. Hung. Acad. Sci.}, 5, 1960.

\bibitem{Fort}
S.~Fortunato.
\newblock Community detection in graphs.
\newblock {\em Physics}, 486:75--174, 2010.

\bibitem{Fowl}
J.~H. Fowler and N.~A. Christakis.
\newblock Dynamic spread of happiness in a large scale network: longitudinal
  analysis over 20 years in the framingham heart study.
\newblock {\em BMJ}, 337, 2008.

\bibitem{Fowl2}
J.~H. Fowler and N.~A. Christakis.
\newblock Estimating peer effects on health in social networks.
\newblock {\em J. Health Econ.}, 27:1400--1405, 2008.

\bibitem{Diego}
D.~Garlaschelli.
\newblock The weighted random graph model.
\newblock {\em New Journal of Physics}, 11, 2009.

\bibitem{Good}
S.~Goodreau, J.~Kitts, and M.~Morris.
\newblock Birds of a feather, or friend of a friend? using exponential random
  graph models to investigate adolescent social networks.
\newblock {\em EPL}, 87, 2009.

\bibitem{Granovetter}
M.~Granovetter.
\newblock The strength of weak ties.
\newblock {\em American Journal of Sociology}, 78:1360--1380, 1973.

\bibitem{Harling}
G.~Harling and J.-P. Onnela.
\newblock Impact of degree truncation on the spread of a contagious process on
  networks.
\newblock 2016.

\bibitem{Kim2}
D.~Kim, A.~Hwong, D.~Stafford, D.~Hughes, A.~O'Malley, J.~Fowler, and
  N.~Christakis.
\newblock Social network targeting to maximise population behaviour change: a
  cluster randomised controlled trial.
\newblock {\em The Lancet}, 386, 2015.

\bibitem{Kim}
D.~Kim, A.~O'Malley, and J.-P. Onnela.
\newblock The social geography of american medicine.
\newblock 2016.

\bibitem{Kumpula}
J.~Kumpula, J.-P. Onnela, J.~Saramaki, K.~Kaski, and J.~Kertesz.
\newblock Emergence of communities in weighted networks.
\newblock {\em Physical Review Letters}, 99, 2007.

\bibitem{Landon}
B.~Landon, N.~Keating, M.~Barnett, J.-P. Onnela, S.~Paul, A.~O'Malley,
  T.~Keegan, and N.~Christakis.
\newblock Variation in patient-sharing networks of physicians across the united
  states.
\newblock {\em JAMA}, 308:265--273, 2012.

\bibitem{Oxford}
K.~Lin.
\newblock Motif counts, clustering coefficients and vertex degrees in models of
  random networks.
\newblock 2007.

\bibitem{Polio}
J.-P. Onnela, B.~Landon, A.~Kahn, D.~Ahmed, H.~Verma, A.~O'Malley, S.~Bahl,
  R.~Sutter, and N.~Christakis.
\newblock Polio vaccine hesitancy in the networks and neighborhoods of
  malegaon, india.
\newblock {\em Social Science and Medicine}, 2016.

\bibitem{Onnela}
J.-P. Onnela, J.~Saramaki, J.~Hyvonen, G.~Szabo, D.~Lazer, K.~Kaski,
  J.~Kertesz, and A.-L. Barabasi.
\newblock Structure and tie strengths in mobile communication networks.
\newblock {\em PNAS}, 104:7332--7336, 2007.

\bibitem{Min}
N.~Papadatos.
\newblock Maximum variance of order statistics.
\newblock {\em Ann. Inst. Statist. Math}, 47:185--193, 1995.

\bibitem{Porter}
M.~Porter, J.-P. Onnela, and P.~J. Mucha.
\newblock Communities in networks.
\newblock {\em Notices of the AMS}, 56:1082 -- 1166, 2009.

\bibitem{Saramaki}
J.~Saramaki, J.-P.~O. M.~Kivela, K.~Kaski, and J.~Kertesz.
\newblock Generalizations of the clustering coefficient to weighted complex
  networks.
\newblock {\em Physical Review E}, 2007.

\bibitem{Sima}
C.~Sima, K.~Panageas, G.~Heller, and D.~Schrag.
\newblock Analytical strategies for characterizing chemotherapy diffusion with
  patient-level population-based data.
\newblock {\em Appl Health Econ Health Policy}, 8:37--51, 2010.

\bibitem{Skellam}
J.~Skellam.
\newblock The frequency distribution of the difference between two poisson
  variates belonging to different populations.
\newblock {\em Journal of the Royal Statistical Society}, 109, 1946.

\bibitem{Staples}
P.~Staples, E.~Ogburn, and J.-P. Onnela.
\newblock Incorporating contact network structure in cluster randomized trials.
\newblock {\em Scientific Reports}, 5, 2015.

\bibitem{Kendall}
A.~Stuart and K.~Ord.
\newblock {\em Kendall's Advanced Theory of Statistics: v.1}.
\newblock Wiley-Blackwell, 1998.

\bibitem{Tore}
O.~Tore.
\newblock Triadic closure in two-mode networks: Redefining the global and local
  clustering coefficients.
\newblock {\em Social Networks}, 35:159--167, 2013.

\bibitem{Valente}
T.~Valente.
\newblock Network models and methods for studying the diffusion of innovations.
\newblock {\em Models and Methods in Social Network Analysis}, pages 98--116,
  2005.

\bibitem{VanderWeele}
T.~VanderWeele.
\newblock Sensitivity analysis for contagion effects in social networks.
\newblock {\em Sociological Methods and Research}, 40:240--255, 2011.

\bibitem{Wang}
J.~Wang, M.~Li, H.~Wang, and Y.~Pan.
\newblock Identification of essential proteins based on edge clustering
  coefficient.
\newblock {\em IEEE/ACM Trans. Comput. Biol. Bioinformatics}, 9(4):1070?1080,
  July 2012.

\bibitem{Faust}
S.~Wasserman and K.~Faust.
\newblock {\em Social network analysis: Methods and applications}.
\newblock Cambridge University Press, 1994.

\bibitem{Watts}
D.~Watts and S.~Strogatz.
\newblock Collective dynamics of `small-world' networks.
\newblock {\em Nature}, 393:440--442, 1998.

\end{thebibliography}

\newpage
\appendix
\section{Approach 1 Mean and Variance Derivations}
\subsection{Original \er Overlap}
\hspace{0.45cm} Edge overlap is considered a random variable with some well-defined but unknown mean and variance (See Eq. \eqref{eq:a}). We first define the distributions of the variables used to define overlap (denoted by uppercase letters) and then proceed to approximate its mean and variance. For each approximation, we assume $n$ is large.
\begin{eqnarray}\label{eq:a} 
O_{ij} = \frac{n_{ij}}{(k_i -1) + (k_j -1) - n_{ij}} \Rightarrow  \frac{N_{ij}}{K_i + K_j - 2 - N_{ij}} =  \frac{N_{ij}}{H_{ij}}
\end{eqnarray}

Suppose we have an \er random graph with $n$ nodes and connection probability $p$. The probability that both $i$ and $j$ are connected to a common neighbor $k$ is equal to $p^2$, and the total number of possible common neighbors is equal to $n-2$. Thus, the distribution of the number of common neighbors, $N_{ij}$, is a binomial random variable with $n-2$ trials and connection probability $p^2$. For large $n$, this can be approximated with a Poisson$(np^2)$ distribution. Similarly, the probability that one node is connected to another is $p$, and each node has a total of $n-1$ other nodes it could connect to. Thus, the degree distribution, $K_i$, is also a binomial random variable with $n-1$ trials and probability $p$. This can also be approximated by a Poisson$(np^2)$ for large $n$. Using the Poisson approximations, the denominator becomes the difference between two Poisson random variables, $H_{ij} = (K_i + K_j - 2)$ and $N_{ij}$, which is a Skellam random variable \cite{Skellam}. Table \ref{table: a} summarizes these distributions.

\begin{table}[htb]
\caption{The distribution, mean and variance for each random variable included in \er overlap.}
\label{table: a}
\begin{minipage}{\textwidth}
\centering
\begin{tabular}{clll}
  Variable & \hspace{0.1cm}Distribution & \hspace{0.1cm}Mean & \hspace{0.1cm}Variance  \\ \hline
  $N_{ij}$ & \hspace{0.1cm}Poisson$(np^2)$ & \hspace{0.1cm}$np^2$ & \hspace{0.1cm}$np^2$ \\
  $K_i, K_j$ & \hspace{0.1cm}Poisson$(2np - 2)$ & \hspace{0.1cm}$2np - 2$ & \hspace{0.1cm}$2np - 2$ \\
  $H_{ij}$ & \hspace{0.1cm}Skellam$(2np - 2 - np^2)$ & \hspace{0.1cm}$2np - 2 - np^2$ & \hspace{0.1cm}$2np - 2 + np^2$ \\
\end{tabular}
\end{minipage}
\end{table}

Edge overlap is the ratio of two random variables and its mean and variance can be approximated using a Taylor series expansion \cite{Kendall, Johnson}. The general form of a first order Taylor series expansion for a function $g(x) = g(x_1, x_2, \dots, x_k)$ about $\theta = (\theta_1, \theta_2, \dots, \theta_k)$ is
\begin{eqnarray}\label{eq:tay}
g(x) = g(\theta) + \sum_{i=1} ^k g_{i}'(\theta)(x_i - \theta_i) + O(n^{-1})
\end{eqnarray}

where $g'(x)$ denotes the derivative of $g(x)$. Here, the function is the ratio of $N_{ij}$ over $H_{ij}$. Define $g(N_{ij}, H_{ij}) = \frac{N_{ij}}{H_{ij}}$ where $H_{ij}$ has no mass at 0. This assumption is assured by the constraints defined in the Methods Section of the paper. Equation \eqref{eq:tay_mean} shows the Taylor series expansion for $g(N_{ij}, H_{ij})$ about the mean, $\theta = (\mathbb{E}(N_{ij}), \mathbb{E}(H_{ij}))$.

\begin{eqnarray}\label{eq:tay_mean}
g(N_{ij}, H_{ij}) &=& g(\theta) + \sum_{i=1} ^2 g_{i}'(\theta)(x_i - \theta_i) + O(n^{-1}) \\[15 pt] \nonumber
&=& g(\theta) + g_{N_{ij}}'(\theta)(N_{ij} - \theta_{N_{ij}}) + g_{H_{ij}}'(\theta)(H_{ij} - \theta_{H_{ij}}) + O(n^{-1}) \\[15 pt] \nonumber
&=& g(\theta) + g_{N_{ij}}'(\theta)(N_{ij} - \mathbb{E}(N_{ij})) + g_{H_{ij}}'(\theta)(H_{ij} - \mathbb{E}(H_{ij})) + O(n^{-1})
\end{eqnarray}

Using the above approximation, the expectation of the ratio, $\mathbb{E}[g(N_{ij}, H_{ij})]$, can be derived as in Eq. \eqref{eq:tay_exp}. 

\begin{eqnarray}\label{eq:tay_exp}
\mathbb{E}[g(N_{ij}, H_{ij})] &=& \mathbb{E}[g(\theta) + g_{N_{ij}}'(\theta)(N_{ij} - \mathbb{E}(N_{ij})) \\[15pt] \nonumber
&+& g_{H_{ij}}'(\theta)(H_{ij} - \mathbb{E}(H_{ij})) + O(n^{-1})] \\[15pt] \nonumber
&=& \mathbb{E}[g(\theta)] + \mathbb{E}[g_{N_{ij}}'(\theta)(N_{ij} - \mathbb{E}(N_{ij}))] +\mathbb{E}[ g_{H_{ij}}'(\theta)(H_{ij} - \mathbb{E}(H_{ij}))]  \\[15pt] \nonumber
&=& \mathbb{E}[g(\theta)] + g_{N_{ij}}'(\theta)\mathbb{E}[N_{ij} - \mathbb{E}(N_{ij})] + g_{H_{ij}}'(\theta)\mathbb{E}[H_{ij} - \mathbb{E}(H_{ij})]  \\[15pt] \nonumber
&=& \mathbb{E}[g(\theta)] + 0 + 0 \approx  g(\mathbb{E}(N_{ij}), \mathbb{E}(H_{ij})) = \frac{\mathbb{E}(N_{ij})}{\mathbb{E}(H_{ij})}\\[15pt]\nonumber
&=& \frac{np^2}{2np - 2 - np^2} \approx \frac{p}{2-p}
\end{eqnarray}


Using the definition of variance and the result that $\mathbb{E}[g(N_{ij}, H_{ij})] \approx g(\theta)$ from Eq. \eqref{eq:tay_exp}, the variance of $g(N_{ij}, H_{ij})$ can be first approximated by Eq. \eqref{eq:var1}.

\begin{eqnarray}\label{eq:var1}
\text{Var}(g(N_{ij},H_{ij})) &=& \mathbb{E}\left\{[g(N_{ij}, H_{ij}) - \mathbb{E}(g(N_{ij}, H_{ij}))]^2  \right\} \\[15pt] \nonumber
&\approx&  \mathbb{E}\left\{[g(N_{ij}, H_{ij}) - g(\theta)]^2  \right\} 
\end{eqnarray}

Using the first order Taylor expansion for $g(N_{ij}, H_{ij})$ from Eq. \eqref{eq:tay_mean}, we have
\begin{eqnarray}\label{eq:var2}
\text{Var}(g(N_{ij},H_{ij})) &\approx& \mathbb{E}\{ [g(\theta) + g_{N_{ij}}'(\theta)(N_{ij} - \mathbb{E}(N_{ij})) \\[15pt] \nonumber
&+& g_{H_{ij}}'(\theta)(H_{ij} - \mathbb{E}(H_{ij})) - g(\theta)]^2\}\\[15pt] \nonumber
&=&\mathbb{E}\left\{ [g_{N_{ij}}'(\theta)(N_{ij} - \mathbb{E}(N_{ij})) + g_{H_{ij}}'(\theta)(H_{ij} - \mathbb{E}(H_{ij}))]^2 \right\}\\[15pt] \nonumber
&=& \mathbb{E}[g'_{N_{ij}^2}(\theta)(N_{ij} - \mathbb{E}(N_{ij}))^2 + g'_{H_{ij}^2}(\theta)(H_{ij} - \mathbb{E}(H_{ij}))^2 \\ [15pt] \nonumber
&+& 2g_{N_{ij}}'(\theta)(N_{ij} - \mathbb{E}(N_{ij}))g_{H_{ij}}'(\theta)(H_{ij} - \mathbb{E}(H_{ij}))]\\[15pt] \nonumber
&=& g'_{N_{ij}^2}(\theta)\text{Var}(N_{ij}) + g'_{H_{ij}^2}(\theta)\text{Var}(H_{ij}) \\[15pt] \nonumber
&+& 2g_{N_{ij}}'(\theta)g_{H_{ij}}'(\theta)\text{Cov}(N_{ij},H_{ij}).
\end{eqnarray}

In this case, $g(N_{ij}, H_{ij}) = \frac{N_{ij}}{H_{ij}}$, $g'_{N_{ij}} = \frac{1}{H_{ij}}$, $g'_{H_{ij}} = \frac{-N_{ij}}{H_{ij}^2}$, and $\theta = (\mathbb{E}(N_{ij}),\mathbb{E}(H_{ij}))$, \\$g'_{N_{ij}}(\theta)g'_{H_{ij}}(\theta) = \frac{-\mathbb{E}(N_{ij})}{\mathbb{E}^3(H_{ij})}$, $g'_{N_{ij}^2}(\theta) = \frac{1}{\mathbb{E}^2(H_{ij})}$, $g'_{H_{ij}^2}(\theta) = \frac{\mathbb{E}^2(N_{ij})}{\mathbb{E}^4(H_{ij})}$. Placing these expressions into \eqref{eq:var2} we have that
\begin{eqnarray}\label{eq:var3}
\text{Var}(g(N_{ij},H_{ij})) &\approx& \frac{\text{Var}(N_{ij})}{\mathbb{E}^2(H_{ij})} + \frac{\mathbb{E}^2(N_{ij})\text{Var}(H_{ij})}{\mathbb{E}^4(H_{ij})} - 2\frac{\text{Cov}(N_{ij}, H_{ij})\mathbb{E}(N_{ij})}{\mathbb{E}^3(H_{ij})} \\[15pt] \nonumber
&=& \frac{\mathbb{E}^2(N_{ij})}{\mathbb{E}^2(H_{ij})}\left[ \frac{\text{Var}(N_{ij})}{\mathbb{E}^2(N_{ij})} + \frac{\text{Var}(H_{ij})}{\mathbb{E}^2(H_{ij})} - 2\frac{\text{Cov}(N_{ij}, H_{ij})}{\mathbb{E}(N_{ij})\mathbb{E}(H_{ij})}  \right] \\[15pt] \nonumber
&=& \frac{np^2}{(2np - 2 - np^2)^2} + \frac{n^2p^4(2np - 2 + np^2)}{(2np - 2 - np^2)^4} - 2\frac{np^2\text{Cov}(N_{ij}, H_{ij})}{(2np - 2 - np^2)^3}.\\[15pt] \nonumber
\end{eqnarray}
Note that Cov$(N_{ij}, H_{ij}) > 0$ since $N_{ij} \not\!\perp\!\!\!\perp H_{ij}$. The value for the covariance could be simulated, but for simplicity we choose to ignore this dependence and include only the first two terms of \eqref{eq:var3} in the variance approximation.

A second order Taylor series expansion can be used as a more precise approximation of the mean. The second order Taylor expansion for the overlap ratio is 
\begin{eqnarray}
g(N_{ij}, H_{ij}) &=& g(\theta) + g_{N_{ij}}'(\theta)(N_{ij} - \theta_{N_{ij}}) + g_{H_{ij}}'(\theta)(H_{ij} - \theta_{H_{ij}}) \\[15pt] \nonumber 
&+&  \frac{1}{2} g_{N_{ij}N_{ij}}''(\theta)(N_{ij} - \theta_{N_{ij}})^2 +  \frac{1}{2}g_{H_{ij}H_{ij}}''(\theta)(H_{ij} - \theta_{H_{ij}})^2 \\[15 pt]\nonumber 
&+&  g_{N_{ij}H_{ij}}''(\theta)(N_{ij} - \theta_{N_{ij}})(H_{ij} - \theta_{H_{ij}})  + O(n^{-1}) \\[15 pt]\nonumber 
&=& g(\theta) + g_{N_{ij}}'(\theta)(N_{ij}- \mathbb{E}(N_{ij})) + g_{H_{ij}}'(\theta)(H_{ij} - \mathbb{E}(H_{ij}))\\[15pt] \nonumber 
&+&  \frac{1}{2}g_{N_{ij}N_{ij}}''(\theta)(N_{ij} - \mathbb{E}(N_{ij}))^2 + \frac{1}{2}g_{H_{ij}H_{ij}}''(\theta)(H_{ij} - \mathbb{E}(H_{ij}))^2 \\[15pt] \nonumber 
&+& g_{N_{ij}H_{ij}}''(\theta)(N_{ij} - \mathbb{E}(N_{ij}))(H_{ij} - \mathbb{E}(H_{ij})) + O(n^{-1}). \nonumber 
\end{eqnarray}

Thus, a better approximation of $E(g(N_{ij}, H_{ij}))$ about $\theta = (\mathbb{E}(N_{ij}), \mathbb{E}(H_{ij}))$ is
\begin{eqnarray}\label{eq:var4}
\mathbb{E}[g(H_{ij}, N_{ij})] &=& \mathbb{E}[g(\theta) + g_{N_{ij}}'(\theta)(N_{ij}- \mathbb{E}(N_{ij})) + g_{H_{ij}}'(\theta)(H_{ij} - \mathbb{E}(H_{ij}))  \\[15pt] \nonumber
&+& \frac{1}{2}g_{N_{ij}N_{ij}}''(\theta)(N_{ij} - \mathbb{E}(N_{ij}))^2 \frac{1}{2}g_{H_{ij}H_{ij}}''(\theta)(H_{ij} - \mathbb{E}(H_{ij}))^2 \\[15 pt] \nonumber
&+& g_{N_{ij}H_{ij}}''(\theta)(N_{ij} - \mathbb{E}(N_{ij}))(H_{ij} - \mathbb{E}(H_{ij})) + O(n^{-1})] \\[15 pt] \nonumber
&=& \mathbb{E}[g(\theta) + \frac{1}{2}\left\{ g_{N_{ij}N_{ij}}''(\theta)\text{Var}(N_{ij}) + g_{H_{ij}H_{ij}}''(\theta)\text{Var}(H_{ij}) \right\}] \\[15pt] \nonumber
&+&  g_{N_{ij}H_{ij}}''(\theta)\text{Cov}(N_{ij},H_{ij}) + O(n^{-1})]. \nonumber
\end{eqnarray}

For $g(N_{ij}, H_{ij}) = \frac{N_{ij}}{H_{ij}}$, $ g_{N_{ij}N_{ij}}'' = 0$, $g_{N_{ij}H_{ij}}'' = \frac{-1}{H_{ij}^2}$, $g_{H_{ij}H_{ij}}'' = \frac{2N_{ij}}{H_{ij}^3}$. Plugging these expressions into \eqref{eq:var4} results in Eq. \eqref{eq:var5}.

\begin{eqnarray}\label{eq:var5}
\mathbb{E}[g(N_{ij}, H_{ij}))] &=& \frac{\mathbb{E}(N_{ij})}{\mathbb{E}(H_{ij})} + \frac{\text{Var}(H_{ij})\mathbb{E}(N_{ij})}{\mathbb{E}^3(H_{ij})} - \frac{\text{Cov}(N_{ij}, H_{ij})}{\mathbb{E}^2(H_{ij})} \\[15pt] \nonumber
&=& \frac{np^2}{2np - 2 - np^2} + \frac{(2np - 2 + np^2)(np^2)}{(2np - 2 - np^2)^3} - \frac{\text{Cov}(N_{ij}, H_{ij})}{(2np - 2 - np^2)^2} \\[15pt] \nonumber
&=& \frac{p}{2 - p} + \frac{(2np - 2 + np^2)(np^2)}{(2np - 2 - np^2)^3} - \frac{\text{Cov}(N_{ij}, H_{ij})}{(2np - 2 - np^2)^2}. \\[15pt] \nonumber
\end{eqnarray}

Again, Cov$(N_{ij}, H_{ij}) > 0$ since $N_{ij} \not\!\perp\!\!\!\perp H_{ij}$. The value for the covariance could be simulated, but for simplicity we chose to ignore this dependence and only include the first two terms of \eqref{eq:var5} in the  approximation of the mean.

Now suppose we introduce weights to the network edges and construct a WRG with $n$ nodes. The weighted \er overlap can be written as in Eq. \eqref{eq:w1}. $N_{ij}$ is again the of the number of common neighbors nodes $i$ and $j$ share, $W_{ij}$ is the weight of the tie between nodes $i$ and $j$ and $S_i(S_j)$ is the strength of node $i(j)$. The ratio is denoted as $V_{ij}$ over $M_{ij}$. We again define the distribution of each of the random variables in the expression and then use the Taylor series expansion approximation outlined in the previous section to derive expressions for the mean and variance of weighted overlap.

\begin{eqnarray}\label{eq:w1}
O^W_{ij} = \frac{\sum^{n_{ij}}_{k=1}(w_{ik} + w_{jk})}{s_i + s_j - 2w_{ij}} \Rightarrow \frac{\sum^{N_{ij}}_{k=1}(W_{ik} + W_{jk})}{S_i + S_j - 2W_{ij}} = \frac{V_{ij}}{M_{ij}}
\end{eqnarray}

For each pair of nodes, an edge is created between them with probability $p$, and a unit weight is added to that edge again with probability $p$ until the first `failure'. This describes a geometric distribution, meaning $W_{ij} \sim$ geometric$(1-p)$. However, to ensure the existence of overlap, we are assuming that the values of all weights are $>0$. Consequently, $W_{ij}$ is a zero-truncated geometric$(1-p)$. The strength of a node is the sum of the weights associated with the edges between that node and all other nodes in the network. Thus, the strength of any node is the sum of $n-1$ geometric random variables, meaning $S_i \sim$ negative binomial$(n-1, 1-p)$ \cite{Diego}. Regardless of the weight of the edge, the probability of an edge existing between nodes $i$ and $j$ is equal to $p$. Therefore, the distribution of $N_{ij}$ is identical to that described in the previous section; a binomial$(n-2, p^2)$. This can again be approximated by a Poisson$(np^2)$ distribution for large $n$.

Focusing on the numerator, $V_{ij}$ is a sum of zero-truncated geometric random variables where the number of variables summed is itself a random variable. More specifically, $V_{ij}$ is a negative binomial random variable with a parameter that depends on the value of $N_{ij}$. We use hierarchical models to calculate the mean (Eq. \eqref{eq:w_mean}) and variance (Eq. \eqref{eq:w_var}) of $V_{ij}$. 

\begin{eqnarray}\label{eq:w_mean}
E[V_{ij}] &=& E[E[V_{ij}|N_{ij}]] = E\left[\frac{2N_{ij}}{(1 - p)}\right]\\[15pt] \nonumber
& = & \frac{2}{(1 - p)}E[N_{ij}] = \frac{2np^2}{(1 - p)} \nonumber
\end{eqnarray}

\begin{eqnarray}\label{eq:w_var}
\text{Var}(V_{ij}) &=& E[\text{Var}(V_{ij}|N_{ij})] + \text{Var}(E[V_{ij}|N_{ij}])\\[15pt]\nonumber
&=& E\left[ \frac{2pN_{ij}}{(1 - p)^2}\right] + \text{Var}\left( \frac{2N_{ij}}{(1 - p)}\right)\\[15pt] \nonumber
&=& \left[ \frac{2p}{(1 - p)^2}\right]E[N_{ij}] + \left[ \frac{2}{(1 - p)}\right]^2\text{Var}(N_{ij})\\[15pt] \nonumber
&=& \frac{2np^2(p+2)}{(1 - p)^2}\nonumber
\end{eqnarray}

The distribution of $M_{ij}$ is more convoluted. In fact, it is unknown, and its mean and variance must be calculated directly (Eqs. \eqref{eq:m_mean} and \eqref{eq:m_var}). Table \ref{table: b} summarizes all of these distributions.

\begin{eqnarray}\label{eq:m_mean}
\mathbb{E}[M_{ij}] &=& \mathbb{E}[S_i] + \mathbb{E}[S_j] - \mathbb{E}[2W_{ij}] \\[15pt] \nonumber
&=& \frac{(n-1)p}{(1-p)} + \frac{(n-1)p}{(1-p)}- \frac{2}{(1-p)} \approx \frac{2np - 2}{(1-p)} 
\end{eqnarray}

\begin{eqnarray}\label{eq:m_var}
\text{Var}(M_{ij}) &=& \text{Var}(S_i) + \text{Var}(S_j) + \text{Var}(2W_{ij})\\[15pt]\nonumber
&=& \frac{(n-1)p}{(1-p)^2} + \frac{(n-1)p}{(1-p)^2}- \frac{4p}{(1-p)^2} = \frac{2np}{(1-p)^2}
\end{eqnarray}

\begin{table}[htb]
\captionsetup{width=\textwidth}
\caption{The distribution, mean and variance for each random variable included in weighted \er overlap.}
\label{table: b}
\begin{minipage}{\textwidth}
\centering
\begin{tabular}{clcc}
  Variable & \hspace{0.1cm}Distribution &  Mean & Variance  \\
\hline
  $W_{ij}$ &\hspace{0.1cm}Zero-truncated Geometric$(1 - p)$ & $\frac{1}{(1 - p)} $ & $\frac{1}{(1 - p)^2}$ \\[15pt]
   $S_{i}, S_j$ &\hspace{0.1cm}\text{Negative Binomial}$(n-1, 1 - p)$ & $\frac{(n- 1)p}{(1 - p)}$ & $\frac{(n- 1)p}{(1 - p)^2}$ \\[15pt]
  $N_{ij}$ &\hspace{0.1cm}Poisson$(np^2)$& $np^2$ & $np^2$ \\[15pt]
$V_{ij}$ & \hspace{0.1cm}Negative Binomial & $\frac{2np^2}{(1 - p)} $ & $\frac{2np^2(p+2)}{(1-p)^2}$ \\[15pt]
  $M_{ij}$ & \hspace{0.1cm}Unknown & $\frac{2np - 2}{(1-p)} $ & $\frac{2np}{(1-p)^2}$ \\
\end{tabular}
\end{minipage}
\end{table}

Now that the mean and variance of the numerator and denominator have been defined, the mean and variance of weighted overlap can be approximated. Define $g(V_{ij}, M_{ij}) = \frac{V_{ij}}{M_{ij}}$. Using the same equations introduced in the previous section, we have 

\begin{eqnarray}
\mathbb{E}[g(V_{ij}, M_{ij})] &\approx&  g(\mathbb{E}(V_{ij}), \mathbb{E}(M_{ij})) = \frac{\mathbb{E}(V_{ij})}{\mathbb{E}(M_{ij})} = \frac{np^2}{np-1} \approx p
\end{eqnarray}

\begin{eqnarray}\label{eq:wvarnew}
\text{Var}(g(V_{ij},M_{ij})) &\approx& \frac{\mathbb{E}^2(V_{ij})}{\mathbb{E}^2(M_{ij})}\left[ \frac{\text{Var}(V_{ij})}{\mathbb{E}^2(V_{ij})} + \frac{\text{Var}(M_{ij})}{\mathbb{E}^2(M_{ij})} - 2\frac{\text{Cov}(V_{ij}, M_{ij})}{\mathbb{E}(V_{ij})\mathbb{E}(M_{ij})}  \right] \\[15pt] \nonumber
&=& p^2 \left[ \frac{p+2}{2np^2} + \frac{1}{2np} - \frac{(1-p)^2\text{Cov}(V_{ij}, M_{ij})}{2np^2(np-1)} \right].
\end{eqnarray}

Note that Cov$(V_{ij}, M_{ij}) > 0$ since $V_{ij} \not\!\perp\!\!\!\perp M_{ij}$. Using the same approach as with the original \er overlap covariance above, regardless of tie weight, we have that 

\begin{eqnarray*}
P(i \text{ is connected to } k \text{ but } j \text{ is not }) &=& p(1-p)\\
P(j \text{ is connected to } k \text{ but } i \text{ is not }) &=& p(1-p)\\
P(i \text{ and } j \text{ are both connected to } k) &=& p^2\\
P(i \text{ nor } j \text{ is connected to } k) &=& (1-p)^2.
\end{eqnarray*}

Incorporating the fact that both $i$ and $j$ each independently connect to a neighbor $k$ according to a geometric($1-p$) distribution, we can calculate the following probabilities for the numerator $V_{ij}$. Note that $k$ only becomes a common neighbor when both $i$ and $j$ connect to $k$. Therefore, the minimum value $V_{ij}$ can take is 2.

\begin{eqnarray*}
P(V_{ij} = 2) &=& p^2(1-p)^2 \\
P(V_{ij} = 3) &=& 2p^3(1-p)^2 \\
P(V_{ij} = 4) &=& 3p^4(1-p)^2 \\
\vdots \\
P(V_{ij} = v) &=& (v-1)p^v(1-p)^2
\end{eqnarray*}

From this we can calculate the expected value of $V_{ij}$ as follows. Note that this coinsides with the derivation above, but we include this alternate derivation in order to frame the derivation of the cov($V_{ij}, M_{ij}$) term below. 
\begin{eqnarray*}
\mathbb{E}[V_{ij}] &=& \sum^\infty_{v=2} v(v-1)p^v(1-p)^2 \\
&=& (1-p)^2\sum^\infty_{v=2}v(v-1)p^v \\
&=& \frac{2p^2}{1-p}
\end{eqnarray*}

We can similarly calculate the following probabilities of $M_{ij}$ and derive its expected value.
\begin{eqnarray*}
P(M_{ij} = 0) &=& (1-p)^2 \\
P(M_{ij} = 1) &=& 2p(1-p)^2 \\
P(M_{ij} = 2) &=& 3p^2(1-p)^2 \\
P(M_{ij} = 3) &=& 4p^3(1-p)^2 \\
\vdots \\
P(M_{ij} = m) &=& (m+1)p^m(1-p)^2
\end{eqnarray*}

\begin{eqnarray*}
\mathbb{E}[M_{ij}] &=& \sum^\infty_{m=0} m(m+1)p^m(1-p)^2 \\
&=& (1-p)^2\sum^\infty_{m=0}m(m+1)p^m \\
&=& \frac{2p^2(p^2 -3p + 3)}{1-p}
\end{eqnarray*}

Now we can derive \text{Cov}$(V_{ij}, M_{ij})$ as follows.

\begin{eqnarray*}
\text{Cov}(V_{ij}, M_{ij}) &=&  \mathbb{E}[(V_{ij} - \mathbb{E}[V_{ij}])(M_{ij} - \mathbb{E}[M_{ij}])] \\ [15pt]
&=& \sum^{\infty}_{m = 0}\sum^{\infty}_{v = 2}\left[P(M_{ij} = m)P(V_{ij} = v)(V_{ij} - \mathbb{E}[V_{ij}])(M_{ij} - \mathbb{E}[M_{ij}]) \right]\\ [15pt]
&=& \sum^{\infty}_{m = 0}P(M_{ij} = m)(M_{ij} - \mathbb{E}[M_{ij}])\sum^{\infty}_{v = 2}P(V_{ij} = v)(V_{ij} - \mathbb{E}[V_{ij}]) \\ [15pt]
&=& \sum^{\infty}_{m = 0}P(M_{ij} = m)(M_{ij} - \mathbb{E}[M_{ij}])\sum^{\infty}_{v = 2} (v-1)p^v(1-p)^2 \left(v - \frac{2p^2}{1-p} \right) \\[15pt]
&=& \sum^{\infty}_{m = 0}P(M_{ij} = m)(M_{ij} - \mathbb{E}[M_{ij}])(1-p)^2\sum^{\infty}_{v = 2}\left [ v(v-1)p^v - (v-1)p^v\frac{2p^2}{1-p} \right]\\ [15pt]
&=& \sum^{\infty}_{m = 0}P(M_{ij} = m)(M_{ij} - \mathbb{E}[M_{ij}])(1-p)^2 (2p^2(1+p))\\ [15pt]
&=& 2p^2(1+p)\sum^{\infty}_{m = 0}(m+1)p^m(1-p)^2 \left(m - \frac{2p^2(p^2-3p+3)}{1-p} \right) \\ [15pt]
&=& 2p^2(1+p)(1-p)^2 \sum^{\infty}_{m = 0}m(m+1)p^m - (m+1)p^m \left(\frac{2p^2(p^2-3p+3)}{1-p} \right) \\ [15pt]
&=& 2p^2(1+p)(1-p)^2\left[\frac{2p - 2p^2(p^2-3p+3)}{(1-p)^3} \right] \\ [15pt]
&=& 4p^3(1+p)\left[\frac{1 - p(p^2-3p+3)}{1-p} \right] \\
\end{eqnarray*}

In a network of $n$ nodes there are a total of $n-2$ possible neighbors both $i$ and $j$ could connect to, meaning we have a total of $n-2$ trials and the resulting covariance term $\text{Cov}(V_{ij}, M_{ij}) = (n-2)4p^3(1+p)\left[\frac{1 - p(p^2-3p+3)}{1-p} \right]$. As with previous derivations, we will assume $n$ is large and approximate $n-2$ with $n$. Plugging this in to equation \eqref{eq:wvarnew} above, we have 
\begin{eqnarray}
\text{Var}(g(V_{ij},M_{ij})) &\approx& \frac{p+1}{n} - \frac{p(1-p^2)(1 - p(p^2-3p+3))}{(np-1)}.
\end{eqnarray}

Again, a second order Taylor series expansion can be used as a more precise approximation of the mean. Using the same equations introduced in the previous section, the second order Taylor approximation for the weighted overlap mean is 

\begin{eqnarray}\label{eq:var6}
\mathbb{E}[g(V_{ij}, M_{ij}))] &=& \frac{\mathbb{E}(V_{ij})}{\mathbb{E}(M_{ij})} + \frac{\text{Var}(M_{ij})\mathbb{E}(V_{ij})}{\mathbb{E}^3(M_{ij})} - \frac{\text{Cov}(V_{ij}, M_{ij})}{\mathbb{E}^2(M_{ij})} \\[15pt] \nonumber
&\approx& p + \frac{n^2p^3}{(np-1)^3} - \frac{(1-p)^2}{4(np-1)^2}n4p^3(1+p)\left[\frac{1 - p(p^2-3p+3)}{1-p} \right]  \\[15pt] \nonumber
&=& p + \frac{n^2p^3}{(np-1)^3} - \frac{np^3(1-p^2)(1 - p(p^2-3p+3))}{(np-1)^2}  \\[15pt] \nonumber
\end{eqnarray}

As with the original overlap derivation above, the second order expansion is technically more accurate, but our results suggest the difference between the first and second order expansions is negligible in practice. We therefore use the first order expansion in the main text due to its simplicity.


\subsection{Directed \er Overlap}

Now suppose we introduce directionality to the network edges and construct a directed random graph with $n$ nodes and connection probability $p$. The directed \er overlap can be written as Eq. \eqref{eq:d1}. $A_{ij}$ is the adjacency matrix value from node $i$ to node $j$. If $A_{ij} = 1$, there is a directed edge from $i$ to $j$. $K_i^{\text{in}}$ and $K_i^{\text{out}}$ denote the in and out-degree distributions of node $i$, respectively. Note that because $K^{in}_i$ and $K^{out}_i$ are identically distributed for each node $i$, min$(k_i^{\text{in}}, k_j^{\text{out}}) =$ min$(k_j^{\text{in}}, k_i^{\text{out}})$, and w.l.o.g., we write their sum as $2\text{min}(K_j^{\text{in}}, K_i^{\text{out}})$. We denote the numerator and denominator using $D_{ij}$ and $C_{ij}$ respectively. Again we define the distribution of each of the random variables in the expression and then use the Taylor series expansion approximation outlined in the previous sections to derive expressions for the mean and variance of directed overlap. However, the derivations are more complicated for the directed version and do not have a closed form solution due to the minimum expressions in the denominator.
\begin{eqnarray}\label{eq:d1}
O^D_{ij} =  \frac{ \sum^{n}_{k = 1} (A_{ik}A_{kj} + A_{jk}A_{ki}) }{\text{min}(k_i^{\text{in}}, k_j^{\text{out}}) + \text{min}(k_j^{\text{in}}, k_i^{\text{out}} ) - 1} \Rightarrow \frac{ D_{ij}}{2\text{min}(K_j^{\text{in}}, K_i^{\text{out}}) - 1 } = \frac{ D_{ij}}{C_{ij}}
\end{eqnarray}

Focusing on the numerator, each of the $A_{ik}A_{kj}$ and $A_{jk}A_{ki}$ terms is equal to one if and only if both adjacency matrix values are equal to 1, which happens with probability $p^2$ since each generation of an edge is independent. Thus, each of the terms is a Bernoulli$(p^2)$ random variable, and the numerator consists of a sum of $2n$ Bernoulli random variables, meaning it is a binomial$(2n, p^2)$ random variable. For large $n$, this can be approximated by a Poisson$(2np^2)$ distribution. 

The denominator includes the minimum of two, identically distributed random variables, $K^{in}_i$ and $K^{out}_i$. Due to the constraint of existence mentioned in Section 3.1 above, the in and out degrees of nodes $i$ and $j$ can not equal 0, making them zero-truncated binomial$(n-1, p)$ random variables. We again approximate this with a zero-truncated Poisson$(np)$ distribution. The distribution of the minimum of two Poisson random variables is unknown. However, an expression for the exact mean (Eq. \eqref{eq:min_mean}) and an upper bound for the variance (Eq. \eqref{eq:min_var}) can be derived \cite{Min}. We denote the minimum of two random variables as $K_{(1)}$ and $K^i_{in}, K^i_{out}$ as simply $K_i$. Table \ref{table: c} summarizes these random variables.

\begin{eqnarray}\label{eq:min_mean}
\mathbb{E}[K_{(1)}] &=& \sum^{(n-1)}_{k=1}P(K_{(1)} \geq k) = \sum^{(n-1)}_{k=1}P(K_1 \geq k)^2\\[15pt]
&=& \sum^{(n-1)}_{k=1}\left[\sum^{(n-1)}_{j=k} P(K_i = j)  \right]^2 = e^{-2np}\sum^{(n-1)}_{k=1}\left[\sum^{(n-1)}_{j=k} \frac{(np)^j}{j!} \right]^2 \nonumber
\end{eqnarray}

\begin{eqnarray}\label{eq:min_var}
\text{Var}(K_{(1)}) = 2\text{Var}(K_i) = \frac{2npe^{np}}{e^{np} - 1} \left[ 1 - \frac{np}{e^{np}-1} \right]
\end{eqnarray}

\begin{table}
\captionsetup{width=\textwidth}
\caption{The distribution, mean and variance for each random variable included in directed \er overlap.}
\label{table: c}
\begin{minipage}{\textwidth}
\centering
\begin{tabular}{llll}
  Variable & \hspace{0.02cm}Distribution &  Mean & Variance  \\
\hline
  $A_{ik}A_{kj}$ &\hspace{0.02cm}Bernoulli($p^2$) & $p^2 $ & $p^2(1-p^2)$ \\
  $D_{ij}$ &\hspace{0.02cm}Poisson($2np^2$) & $2np^2 $ & $2np^2$ \\
   $K_i^{\text{in}}, K_i^{\text{out}}$ &\hspace{0.02cm}Zero-truncated Poisson$(np)$ & $\frac{npe^{np}}{e^{np}-1}$ & $\frac{npe^{np}}{e^{np} - 1} \left[ 1 - \frac{np}{e^{np}-1} \right]$ \\
  $K_{(1)}$ &\hspace{0.02cm}Unknown & $e^{-2np}\sum^{(n-1)}_{k=1}\left[\sum^{(n-1)}_{j=k} \frac{(np)^j}{j!} \right]^2$ & $\frac{2npe^{np}}{e^{np} - 1} \left[ 1 - \frac{np}{e^{np}-1} \right]$ \\
  $C_{ij}$ &\hspace{0.02cm}Unknown & $ 2e^{-2np}\sum^{(n-1)}_{k=1}\left[\sum^{(n-1)}_{j=k} \frac{(np)^j}{j!} \right]^2 - 1$ & $\frac{8npe^{np}}{e^{np} - 1} \left[ 1 - \frac{np}{e^{np}-1} \right]$ \\
\end{tabular}
\end{minipage}
\end{table}

Now that the mean and variance of the numerator and denominator have been defined, the mean and variance of directed overlap can be approximated. Define $g(D_{ij}, C_{ij}) = \frac{D_{ij}}{C_{ij}}$. Using the same equations introduced in the previous section, we have 

\begin{eqnarray}
\mathbb{E}[g(D_{ij}, C_{ij})] &\approx&  g(\mathbb{E}(D_{ij}), \mathbb{E}(C_{ij})) \\[15pt]\nonumber
&=& \frac{\mathbb{E}(D_{ij})}{\mathbb{E}(C_{ij})}= \frac{np^2}{e^{-2np}\sum^{(n-1)}_{k=1}\left[\sum^{(n-1)}_{j=k} \frac{(np)^j}{j!} \right]^2 - 0.5}
\end{eqnarray}

\begin{eqnarray}
\text{Var}(g(D_{ij},C_{ij})) &\approx& \frac{\mathbb{E}^2(D_{ij})}{\mathbb{E}^2(C_{ij})}\left[ \frac{\text{Var}(D_{ij})}{\mathbb{E}^2(D_{ij})} + \frac{\text{Var}(C_{ij})}{\mathbb{E}^2(C_{ij})} - 2\frac{\text{Cov}(D_{ij}, C_{ij})}{\mathbb{E}(D_{ij})\mathbb{E}(C_{ij})}  \right] \\[15pt]\nonumber
&=& \frac{2n^2p^4}{(2e^{-2np}\sum^{(n-1)}_{k=1}\left[\sum^{(n-1)}_{j=k} \frac{(np)^j}{j!} \right]^2 - 1)^2} \\[15pt] \nonumber
&+& \frac{\frac{32n^3p^5e^{np}}{e^{np} - 1} \left[ 1 - \frac{np}{e^{np}-1}\right]}{(2e^{-2np}\sum^{(n-1)}_{k=1}\left[\sum^{(n-1)}_{j=k} \frac{(np)^j}{j!} \right]^2 - 1)^2}\\[15pt]\nonumber
& -& \frac{4np^2\text{Cov}(D_{ij}, C_{ij})}{(2e^{-2np}\sum^{(n-1)}_{k=1}\left[\sum^{(n-1)}_{j=k} \frac{(np)^j}{j!} \right]^2 - 1)^3}. 
\end{eqnarray}

Note that Cov$(D_{ij}, C_{ij}) > 0$ since $D_{ij} \not\!\perp\!\!\!\perp C_{ij}$. The value for the covariance could be simulated, but for simplicity we ignore this dependence and do not include the covariance term in the final approximation.

Again, a second order Taylor series expansion can be used as a more precise approximation of the mean. Using the same equations introduced in the previous section, the second order Taylor approximation for the directed overlap mean is 

\begin{eqnarray}\label{eq:var7}
\mathbb{E}[g(D_{ij}, C_{ij}))] &=& \frac{\mathbb{E}(D_{ij})}{\mathbb{E}(C_{ij})} + \frac{\text{Var}(C_{ij})\mathbb{E}(D_{ij})}{\mathbb{E}^3(C_{ij})} - \frac{\text{Cov}(D_{ij}, C_{ij})}{\mathbb{E}^2(C_{ij})} \\[15pt] \nonumber
&=& \frac{np^2}{e^{-2np}\sum^{(n-1)}_{k=1}\left[\sum^{(n-1)}_{j=k} \frac{(np)^j}{j!} \right]^2 - 0.5} \\[15pt] \nonumber
&+& \frac{\frac{16n^2p^3e^{np}}{e^{np} - 1} \left[ 1 - \frac{np}{e^{np}-1} \right]}{(2e^{-2np}\sum^{(n-1)}_{k=1}\left[\sum^{(n-1)}_{j=k} \frac{(np)^j}{j!} \right]^2 - 1)^3} \\[15pt] \nonumber
&-&  \frac{\text{Cov}(D_{ij}, C_{ij})}{(2e^{-2np}\sum^{(n-1)}_{k=1}\left[\sum^{(n-1)}_{j=k} \frac{(np)^j}{j!} \right]^2 - 1)^2}.
\end{eqnarray}

Again, Cov$(D_{ij}, C_{ij}) > 0$ since $D_{ij} \not\!\perp\!\!\!\perp C_{ij}$. The value for the covariance could be simulated, but for simplicity we chose to ignore this dependence and only include the first two terms of Eq. \eqref{eq:var7} in the  approximation of the mean.


\section{Approach 2 Mean and Variance Derivations}

\subsection{Original \er Overlap}
\hspace{0.45cm}Again, suppose we have an \er random graph with $n$ nodes and connection probability $p$. Edge overlap is again viewed as a random variable with the same distributions for the numerator and denominator defined in the first approach described in Section A.3. The expectation of the denominator is equal to $(2np - 2 - np^2)$, and we can rewrite $O_{ij}$ as Eq. \eqref{eq:a2a}.  
\begin{eqnarray}\label{eq:a2a}
O_{ij} = \frac{N_{ij}}{H_{ij}} \approx \frac{N_{ij}}{\mathbb{E}[H_{ij}]} = \frac{1}{2np-2-np^2}N_{ij}
\end{eqnarray}

The distribution of overlap is now a scaled version of the distribution of $N_{ij}$, making it a scaled Poisson$(np^2)$ random variable and its mean (Eq. \eqref{eq:a2_mean}) and variance (Eq. \eqref{eq:a2_var}) can be easily derived.
\begin{eqnarray}\label{eq:a2_mean}
\mathbb{E}[O_{ij}] = \frac{1}{2np-2-np^2}\mathbb{E}[N_{ij}]= \frac{np^2}{2np-2-np^2} \approx \frac{p}{2-p}
\end{eqnarray}

\begin{eqnarray}\label{eq:a2_var}
\text{Var}(O_{ij})  = \frac{1}{(2np-2-np^2)^2}\text{Var}(N_{ij})= \frac{np^2}{(2np-2-np^2)^2} 
\end{eqnarray}

Note that the mean is equivalent to the mean derived in the first approach while the variance is equal to the first term of the variance derived in the first approach. There is no second order approximation in this case since the approximation is not based on a Taylor expansion.

\subsection{Weighted \er Overlap}

Now suppose we have a WRG with $n$ nodes and connection probability $p$, and weighted overlap is again viewed as a random variable with the same distributions for the numerator and denominator defined in Section A.2. The expectation of the denominator is equal to $\frac{2(n-1)p-2}{(1-p)}$, and we can rewrite $O^W_{ij}$ as Eq. \eqref{eq:a2b}. 
\begin{eqnarray}\label{eq:a2b}
O^W_{ij} =  \frac{V_{ij}}{M_{ij}} \approx \frac{V_{ij}}{\mathbb{E}[M_{ij}]} = \frac{(1-p)}{2np-2}V_{ij}
\end{eqnarray}

The distribution of weighted overlap is now a scaled version of the distribution of $V_{ij}$, making it a scaled Compound Poisson random variable. The mean (Eq.  \eqref{eq:cp_mean}) and variance (Eq.  \eqref{eq:cp_var}) are now easily derived.
\begin{eqnarray}\label{eq:cp_mean}
\mathbb{E}[O^W_{ij}] = \frac{(1-p)}{2np-2}\mathbb{E}[V_{ij}]= \frac{(1-p)}{2np-2}\left( \frac{2np^2}{1-p}\right) \approx p
\end{eqnarray}

\begin{eqnarray}\label{eq:cp_var}
\text{Var}(O^W_{ij} ) &=& \frac{(1-p)^2}{(2np-2)^2}\text{Var}(V_{ij})\\[15pt]\nonumber
&=& \frac{(1-p)^2}{(2np-2)^2}\left( \frac{2np^2(p+2)}{(1-p)^2}\right) \\[15pt]\nonumber
&\approx& \frac{np^2(p+2)}{2(np-1)^2}
\end{eqnarray}

Note that the mean is equivalent to the mean derived in the first approach while the variance is equal to the first term of the variance derived in the first approach. As above, there is no second order approximation in this case since the approximation is not based on a Taylor expansion.

\subsection{Directed \er Overlap}

Now suppose we have a directed \er random graph with $n$ nodes and connection probability $p$, and directed overlap is again viewed as a random variable with the same distributions for the numerator and denominator defined in Section A.3. The expectation of the denominator is equal to $2e^{-2np}\sum^{(n-1)}_{k=1}\left[\sum^{(n-1)}_{j=k} \frac{(np)^j}{j!} \right]^2 - 1$, and we can rewrite $O^D_{ij}$ as Eq. \eqref{eq:a2c}. 
\begin{eqnarray}\label{eq:a2c}
O^D_{ij} =  \frac{D_{ij}}{C_{ij}} \approx \frac{D_{ij}}{\mathbb{E}[C_{ij}]} = \frac{1}{2e^{-2np}\sum^{(n-1)}_{k=1}\left[\sum^{(n-1)}_{j=k} \frac{(np)^j}{j!} \right]^2 - 1}D_{ij}
\end{eqnarray}

The distribution of directed overlap is now a scaled version of the distribution of $D_{ij}$, making it a scaled Poisson$(2np^2)$ random variable. The mean (Eq. \eqref{eq:a2d_mean}) and variance (Eq. \eqref{eq:a2d_var}) are now easily derived.
\begin{eqnarray}\label{eq:a2d_mean}
\mathbb{E}[O^D_{ij}] &=& \frac{1}{2e^{-2np}\sum^{(n-1)}_{k=1}\left[\sum^{(n-1)}_{j=k} \frac{(np)^j}{j!} \right]^2 - 1}\mathbb{E}[D_{ij}]\\[15pt]\nonumber
& =& \frac{np^2}{e^{-2np}\sum^{(n-1)}_{k=1}\left[\sum^{(n-1)}_{j=k} \frac{(np)^j}{j!} \right]^2 - 0.5}
\end{eqnarray}

\begin{eqnarray}\label{eq:a2d_var}
\text{Var}(O^D_{ij} ) & = & \frac{1}{(2e^{-2np}\sum^{(n-1)}_{k=1}\left[\sum^{(n-1)}_{j=k} \frac{(np)^j}{j!} \right]^2 - 1)^2}\text{Var}(D_{ij})\\[15pt]\nonumber
&=& \frac{2np^2}{(2e^{-2np}\sum^{(n-1)}_{k=1}\left[\sum^{(n-1)}_{j=k} \frac{(np)^j}{j!} \right]^2 - 1)^2}
\end{eqnarray}

Note that the mean is equivalent to the mean derived in the first approach while the variance is equal to the first term of the variance derived in the first approach. As above, there is no second order approximation in this case since the approximation is not based on a Taylor expansion.

\end{document}